\documentclass[preprint,review,12pt]{elsarticle}

\usepackage{graphicx}
\usepackage{amssymb}
\usepackage{amsfonts}
\usepackage{amsmath}

\usepackage{rotating}

\usepackage{lineno}

\journal{Icarus}

\begin{document}

\begin{frontmatter}


\title{The SuperCam Infrared Spectrometer for the Perseverance Rover of the Mars2020 mission}



\author[1]{Thierry Fouchet}
\author[1]{Jean-Michel Reess}
\author[2]{Franck Montmessin}
\author[2]{Rafik Hassen-Khodja}
\author[1]{Napoléon Nguyen-Tuong}
\author[2]{Olivier Humeau}
\author[1]{Sophie Jacquinod}
\author[2]{Laurent Lapauw}
\author[1]{Jérôme Parisot}
\author[1]{Marion Bonafous}
\author[1]{Pernelle Bernardi}
\author[1]{Frédéric Chapron}
\author[1]{Alexandre Jeanneau}
\author[1]{Claude Collin}
\author[1]{Didier Zeganadin}
\author[1]{Patricia Nibert}
\author[2]{Sadok Abbaki}
\author[2]{Christophe Montaron}
\author[1]{Cyrille Blanchard}
\author[1]{Vartan Arslanyan}
\author[1]{Ourdya Achelhi}
\author[1]{Claudine Colon}
\author[1,3]{Clément Royer}
\author[3]{Vincent Hamm}
\author[3]{Mehdi Beuzit}
\author[3]{François Poulet}
\author[3]{Cédric Pilorget}
\author[1]{Lucia Mandon}
\author[4]{Olivier Forni}
\author[4]{Agnès Cousin}
\author[4]{Olivier Gasnault}
\author[4]{Paolo Pilleri}
\author[4]{Bruno Dubois}
\author[5]{Cathy Quantin}
\author[6]{Pierre Beck}
\author[7]{Olivier Beyssac}
\author[8]{Stéphane Le Mouélic}
\author[9]{Jeffrey R.\ Johnsson}
\author[10]{Timothy H.\ McConnochie}
\author[4]{Sylvestre Maurice}
\author[11]{Roger C.\ Wiens}

\address[1]{LESIA, Observatoire de Paris, Université PSL, CNRS, Sorbonne Université, Université de Paris, 5 place Jules Janssen, 92195 Meudon, France}

\address[2]{Laboratoire Atmosphères, Milieux, Observations Spatiales (LATMOS), UVSQ Université Paris-Saclay, Sorbonne Université, CNES, Paris, France}

\address[3]{Université Paris-Saclay
, CNRS, Institut d’Astrophysique Spatiale, Bâtiment 121, 91405 Orsay, France}

\address[4]{Institut de Recherche en Astrophysique et Planétologie, Université de Toulouse, UPS, CNRS, CNES, Toulouse, Francee}

\address[5]{Univ. Lyon, Univ. Lyon 1, ENSL, CNRS, LGL-TPE, Villeurbanne, 69007, Lyon, France}

\address[6]{Université Grenoble Alpes, CNRS, Institut de Planétologie et d’Astrophysique de Grenoble (IPAG), UMR 5274, Grenoble F-38041, France}

\address[7]{Institut de Minéralogie, de Physique des Matériaux et de Cosmochimie, Museum National d’Histoire Naturelle, CNRS, Sorbonne Université, Paris, France}

\address[8]{Laboratoire de Planétologie et Géodynamique, Université de Nantes, Université d’Angers, CNRS UMR 6112, Nantes, France}

\address[9]{Johns Hopkins University Applied Physics Laboratory, Laurel, MD 20723-6005, United States}

\address[10]{Department of Astronomy, University of Maryland, College Park, MD 20742, United States}

\address[11]{Los Alamos National Laboratory, Los Alamos, New Mexico, USA}
\begin{abstract}
We present the Infrared spectrometer of SuperCam Instrument Suite that enables the Mars 2020 Perseverance Rover to study remotely the Martian mineralogy within the Jezero crater. The SuperCam IR spectrometer is designed to acquire spectra in the 1.3--2.6 $\mu$m domain at a spectral resolution ranging from 5 to 20~nm. The field-of-view of 1.15 mrad, is coaligned with the boresights of the other remote-sensing techniques provided by SuperCam: laser-induced breakdown spectroscopy, remote time-resolved Raman and luminescence spectroscopies, and visible reflectance spectroscopy, and micro-imaging. The IR spectra can be acquired from the robotic-arm workspace to long-distances, in order to explore the mineralogical diversity of the Jezero crater, guide the Perseverance Rover in its sampling task, and to document the samples' environment. We present the design, the performance, the radiometric calibration, and the anticipated operations at the surface of Mars.

\end{abstract}


\end{frontmatter}

\tableofcontents


\section{Introduction}

    The NASA Mars2020 mission and its Perseverance rover will expand the array of assets available for in situ exploration of the Martian geology and climate with an infrared passive spectrometer as part of the SuperCam instrument suite. Up to the Mars 2020 mission, in situ experiments were mostly active instrumentation. The Sojourner and the MER rovers carried an alpha-proton X-ray spectrometer (APXS) \citep{Rieder1997, Rieder2003,Bruckner2003}, supplemented by a Mössbauer spectrometer on board the MER rovers \citep{Squyres2003}, while the Curiosity Rover of the Mars Science Laboratory mission featured a laser-induced breakdown spectroscopy technique (LIBS) provided by the ChemCam instrument \citep{Wiens2012,Maurice2012}, a powder X-ray diffraction and X-ray fluorescence instrument (CheMin) \citep{Blake2012}, a neutron spectrometer (DAN) \citep{Mitrofanov2012}, a gas chromatograph and mass spectrometer (SAM) \citep{Mahaffy2012}, along with a classical APXS spectrometer \citep{Campbell2012}. Passive remote sensing instrumentation was also part of the rover payloads, in the form of multispectral panoramic camera on the MER rovers \citep{Bell2003}, and the Curiosity rover \citep{Bell2012}, or of highly-resolved imaging systems like the RMI camera on ChemCam \citep{Maurice2012} or the MAHLI camera on Curiosity's robotic arm \citep{Edgett2012}. Among all these active instruments or cameras, the sole infrared spectrometer that has ever operated at the surface of Mars was the Miniature Thermal Emission Spectrometer (mini-TES) \citep{Christensen2003a} covering the 5--29 $\mu$m spectal range (340 to 2000 cm$^{-1}$) with a spectral sample interval of 10~cm$^{-1}$.

    In contrast, orbital platforms made an extensive use of solar-reflected infrared spectroscopy and of thermal emission spectroscopy. In the solar-reflected infrared, the Imaging Spectrometer (ISM) on board the Phobos mission was the first to provide hyperspectral images of Mars \citep{Bibring1989}, followed by the Observatoire pour la Minéralogie, l'Eau, les Glaces et l'Activité (OMEGA) on board the MarsExpress mission \citep{Bibring2004}, and the Compact Reconnaissance Imaging Spectrometer (CRISM) on the Mars Reconnaissance Orbiter (MRO) \citep{Murchie2007}. The thermal infrared range was explored much earlier, starting with the pioneer Infrared Spectroscopy Experiment on Mariner 9 \citep{Conrath1973}, which long preceded a collection of thermal spectrometers in the early 2000s: the Thermal Emission Spectrometer (TES) on the Mars Global Surveyor mission \citep{Christensen1992}, the Thermal Emission Imaging System (THEMIS) on Mars Odyssey \citep{Christensen2003b}, and the Mars Climate Sounder instrument (MCS) on MRO \citep{McCleese2007}. The Martian atmosphere was also probed with the stellar and solar occultation technique in the infrared by SPICAM on Mars Express \citep{Bertaux2000,Korablev2006}, NOMAD and ACS on the ExoMars Trace Gas Orbiter \citep{Vandaele2018,Korablev2018}.

    Thanks to all these experiments, orbital infrared spectroscopy proved extremely successful in determining the mineralogical composition of the Martian crust as well as mapping the minerals that trace the weathering and alteration processes throughout the Martian history, and that are keys in determining the evolution of the Martian environment through time \citep{Bibring2006,Poulet2009,Ehlmann2014}. In fact, infrared spectroscopy was instrumental in the processes that led to the selection of the landing sites for the Opportunity, Curiosity, Perseverance, and Rosalind Franklin rovers. The wide mineralogical diversity of the landing sites, the association of minerals with distinctive morphological units, the presence of minerals tracing weathering and alteration processes by liquid water, and the biological potential of the detected minerals were fundamental drivers for the sites selection. This intensive use of infrared sounding to select and investigate the landing sites resulted in the paradoxical situation in which the sites are better explored in the infrared from orbital platforms than from in-situ rovers. The use of different instrumental techniques on satellites and rovers led to difficulties in reconciling the results obtained by orbital and in-situ exploration, and in drawing a consistent picture of the landing sites mineralogy and geology. For all these reasons, the Mars2020 Science Definition Team recommended a solar-reflected infrared instrument as part of the Perseverance rover payload \citep{Mustard2013}.

    The infrared spectrometer of the SuperCam instrument suite has been designed to fulfill this need for in-situ mineralogical investigation in the near-infrared, in particular in the very diagnostic 1.3--2.6~$\mu$m spectral region. It will play a key role in achieving the four scientific objectives of the Mars2020 mission \citep{Farley2020}: i) the investigation of the mineralogy and geology of the Jezero crater as representative of the ancient Martian environment, ii) the assessment of the habitability of this ancient environment, iii) the identification of rocks with a high potential of preserving biosignatures, iv) and the study of the current environmental Martian conditions in the preparation for human exploration. SuperCam infrared spectrometer, along with the visible/near-infrared spectrometers \citep{Wiens2020,Maurice2020}, will serve as the sole spectroscopic remote sensing instrument of the Mars2020 payload and will play a leading role in the determination of the mineralogy of the terrains explored by the rover. Data acquired by the spectrometers will be instrumental in the selection of targets explored by contact instruments for a refined characterization, and eventually in the choice of the samples that will be collected for a future return to Earth.

    This paper provides a detailed description of the SuperCam infrared spectrometer, its scientific objectives, its design and implementation, its measured performances, and its operation procedures at the surface of Mars. This paper is meant to serve as the reference paper for future use and publication of the SuperCam infrared spectrometer dataset, either from the M2020 team or archive users. The paper is divided into four sections. Section~\ref{Scientificbackground} presents previous infrared investigations of the Martian surface and atmosphere with an emphasis on Jezero crater and its Nilli Fossae context. Section~\ref{Scientificobjectives} presents the specific scientific objectives of the infrared spectrometer. Section~\ref{Instrumentdescription} presents how these scientific objectives were translated into measurement performances and how the SuperCam infrared spectrometer was designed to meet these performances within the available resources. Finally, section~\ref{InstrumentTest} describes how the instrument was tested and calibrated, and how it will be operated on Mars.
    
\section{Scientific background}
\label{Scientificbackground}

    \subsection{How infrared spectrometers have changed our views on Mars mineralogy and history}
    \label{SecIRHistory}
    
        The near-infrared spectral range has long been recognized as very favorable for identifying diagnostic features of rock-forming minerals, although dust obscuring surface signatures was seen as a potential problem. Specifically, most hydrous and anhydrous silicates, water-bearing minerals, carbonates, sulfates can be detected in the near-infrared, solar-reflected, range. Of special interest is the detection of the secondary minerals, which can provide key clues for deciphering the geochemical conditions of their formation, and the history of the water-related alteration. Numerous investigations done by imaging spectrometers, OMEGA/Mars express and CRISM/MRO in the 1.0--2.5 $\mu$m, have revealed a high degree of mineralogical diversity down to sub-kilometer scale \citep{Bibring2006,Murchie2009}. In particular, the detection and the mapping of at least four classes of altered minerals (phyllosilicates, sulfates, carbonates, hydrated silica) indicates that Mars has likely hosted conditions that favored liquid water to remain stable over long periods of time \citep{Gendrin2005,Poulet2005,Ehlmann2008, Milliken2008, Carter2013b}. Rocks containing these minerals are considered to be among the best targets in which in situ investigations by Mars2020 have chances of finding potential biosignatures at microscopic scale.
    
        The widespread occurrence of these altered products and the diversity of settings in which they occur point to a rich and complex history of water-rock interaction that spanned at least hundreds of millions of years starting with the formation of phyllosilicates. These phases (also referred as to clay minerals) are mainly found in ancient Noachian-aged terrains with various environmental settings \citep{Bibring2006,Murchie2009}. Some ancient terrains with clay minerals are hypothesized to form via subsurface water-rock reactions, which did not require a warm, wet Mars for extended periods. These indicator minerals of subsurface waters are hydrothermal/low-grade metamorphic minerals like prehnite and aqueous mineral assemblages with compositions isochemical to basalt \citep{Ehlmann2019}. In contrast, numerous occurrences of stratified terrains with Al-phyllosilicates atop Fe/Mg-phyllosilicates have been found in several regions of Mars, at times with contiguous stratigraphy extending regionally over hundreds or thousands of kilometers \citep{Loizeau2012,Carter2015}. These settings sometimes mixed with sulfates were preferentially formed in variable aqueous environments on the surface of Mars including weathering sequences in a warm, wet early Mars climate  \citep{Bishop2018}. Clay-bearing strata associated with deltas (Jezero but also in younger deltas such as Eberswalde), and other fluvial features, are also strong evidence for sediment surface transport of previously altered basaltic material \citep[e.g.\ ][]{Poulet2014}.
    
        The relative paucity of phyllosilicates in younger Hesperian terrains has led to the hypothesis that the planet transitioned to a drier and more acidic environment as evidenced by the occurrences of sulfate-bearing deposits found primarily in these Hesperian terrains. Rover-scale observations at Meridiani allow speculation that sulfate assemblages observed elsewhere on Mars, such as those in Valles Marineris, may record similar processes. Hence, the large deposits of sulfate-rich sediments formed during the Hesperian could have been the result of widespread (and transient) lacustrine environments \citep{Andrews-Hanna2011}. But the sulfates could be also the legacy of colder, icier surface environments. The origin of sulfates in the north polar region of Mars detected by OMEGA/Mars express is likely related to the polar ice \citep{Langevin2005}. It has been demonstrated that acidic fluids in ice deposits can react with olivine minerals to form sulfates. Consequently, sublimation residue of ancient ice deposits should resemble fine grained, layered, sedimentary deposits with substantial sulfate mineral content as observed on Mars \citep{Masse2012,Michalski2013}.
    
        In contrast to sulfate-bearing and clay-bearing deposits, the hydrated silica mineral detections occur at much smaller spatial scales and are only observable at resolutions of tens to hundreds of meters \citep{Murchie2007,Milliken2008,Ehlmann2009}. Silica-enriched sediments are considered important in terms of biosignature preservation, subsurface hydrothermal activity is commonly evoked to explain their formation. Moreover, hydrated silica are present more frequently in Martian delta deposits than in their terrestrial counterparts, which may point to peculiar environmental conditions prevailing at the epoch of Martian deltas formation \citep{Pan2020}.
    
        The most widespread Noachian carbonate-bearing terrains that could provide a lithologic record of the carbon cycle and atmosphere on early Mars is found in terrains surrounding Jezero crater \citep{Ehlmann2009,Goudge2017}. It is proposed that the carbonate signature in this region is a cement precipitated from carbonate-saturated fluids, possibly as a result of the dissolution of carbonate in the watershed. Interaction with atmospheric CO$_2$ leading to the formation of marginal carbonates could have also occurred during lacustrine activity \citep{Horgan2020}. 
    
        The NIR imaging spectrometers also provided clues about bulk mineralogy of the Martian crust and past volcanic edifices. Olivine is detected in ejecta and/or intra-crater dunes associated with large craters ($> 20$~km) of the northern plains \citep{Ody2013}. These craters are large enough to have excavated the oldest Noachian crust buried under the northern plains demonstrating that this oldest Noachian crust was olivine-bearing at least in some locations in the northern hemisphere. Olivine is also associated with Noachian buttes in the vicinity of the Hellas basin outside its terraces \citep{Ody2013}. This suggests that the oldest Noachian crust was olivine-bearing in some locations of the southern hemisphere. Orbital NIR data also indicate occurrences of felsic rocks at localities distant by hundreds of km, isolated within spatially dominant mafic/basaltic Noachian terrains \citep{Carter2013a}. They could represent a buried ancient evolved crust or more likely localized plutons. The Hesperian mare-style, flood volcanism observed in various edifices such as Syrtis Major Planum and Hesperia Planum is dominated by basaltic compositions with pyroxene, plagioclase and olivine compositions \citep{Ody2012,Riu2019}. High-Calcium Pyroxene (HCP) is the predominant pyroxene type in Hesperian volcanic plains \citep{Mustard2005,Poulet2009}, with the lowest relative Low Calcium Pyroxene (LCP) values in Syrtis Major \citep{Mustard2005,Riu2019}. Compositional difference in pyroxene type has been observed in older terrains, which has been interpreted as a progressive cooling of the mantle \citep{Baratoux2013}. If Mars surface is essentially basaltic, its crust and volcanic components are thus far from being uniform in composition as revealed by NIR data (along with numerous in situ observations) and point to a complex picture of the Noachian and Hesperian magmatism with unexpected petrological diversity. By contrast, most Amazonian volcanic landforms on Mars are located in dusty regions, limiting NIR spectroscopic observations to few dust-free windows. CRISM data on rims of fresh craters probing below dust have recently been used to provide more input into in Tharsis and Elysium volcanic areas. Results show a basaltic composition with no major changes during the Amazonian period, but a significant contribution of LCP \citep{Viviano-Beck2017}.
    
    \subsection{The diversity of infrared signatures in Jezero}
    
\begin{figure*}
            \centering
            \includegraphics[width=15cm,clip=true, trim= 50mm 0 50mm 0]{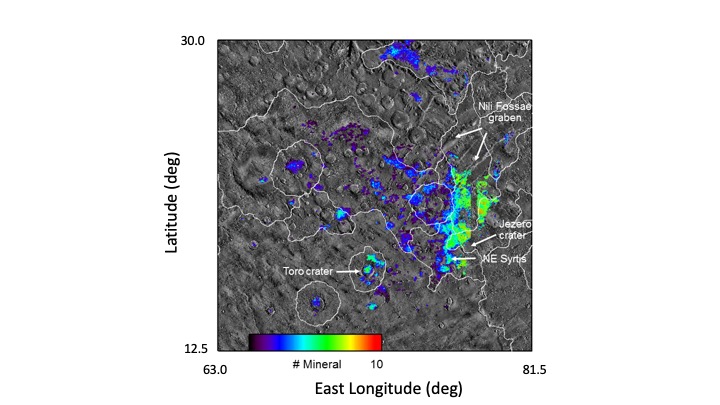}
            \caption{Maps of the number of different minerals identified in Nilli Fossae region from using the OMEGA and CRISM specto-imaging instruments.}
            \label{FigNilliFossae}
        \end{figure*}

        Since the first global spectroscopic survey of the Martian surface in the infrared, several authors have reported that Nili fossae area is a specific region of interest. With one of the lowest dust content of the surface of Mars \citep{Koeppen2008,Ody2012},  the region has revealed a large diversity of minerals detected from orbit by IR spectrometers (Fig.~\ref{FigNilliFossae}). TES/MGS was the first to highlight the most widespread deposit enriched in Olivine in Nili Fossae region \citep{Koeppen2008}, confirmed next by OMEGA/MEX hyperspectral images \citep{Mustard2007,Combe2008,Ody2012,Poulet2007,Mangold2007}. Nili fossae is also the region where the first carbonate signatures have been detected on Mars \citep{Ehlmann2008} and where the most diverse hydrated mineralogy has been observed from phyllosilicates to sulfates \citep{Ehlmann2009}.
        
        Jezero crater, the landing site of Mars2020 mission is located in the Nili fossae region, on the north west margin of Isidis impact basin. Jezero crater is incised by inlet valleys draining a large watershed area characterized by a diverse mineralogy \citep{Goudge2015}. The watershed area includes:  Fe/Mg-smectites, Mg-rich carbonates, olivine, and both calcium rich and calcium poor pyroxenes \citep{Ehlmann2008,Ehlmann2009,Goudge2015,Mangold2007,Mustard2007,Mustard2009}, which all exhibit diagnostic vibrational and crystal field absorptions in SuperCam infrared spectrometer wavelength range. Within Jezero crater, three main geological units with distinct mineralogy are observed \citep{Goudge2015}:  the olivine/carbonate rich unit, a capping floor unit enriched in mafic minerals with no hydrated minerals detected so far, and the deltaic deposit. The deltaic fan is dominated by Fe/Mg-bearing smectites \citep{Goudge2015} and low calcium pyroxenes \citep{Horgan2020} with isolated exposures of Mg-rich carbonate and olivine \citep{Goudge2015}. The diversity of the Jezero crater mineralogy has been enriched by recent studies. \citet{Horgan2020} identified carbonates spectrally distinct from the surrounding carbonates in the margin of Jezero crater, which they interpreted as distal lacustrine carbonates, while \citet{Tarnas2019} detected hydrated silica in marginal deposits that may be correlated to a paleo-lacustrine activity. These two recent detections are important for the search for biosignatures, as marginal carbonates are often biologically mediated on Earth \citep{Capezzuoli2014}, and as opaline silica have a high biosignatures preservation potential \citep{McMahon2018}.
	    
\begin{figure*}
	        \centering
	        \includegraphics[width=15cm,clip=true,trim=0 0 0 270mm]{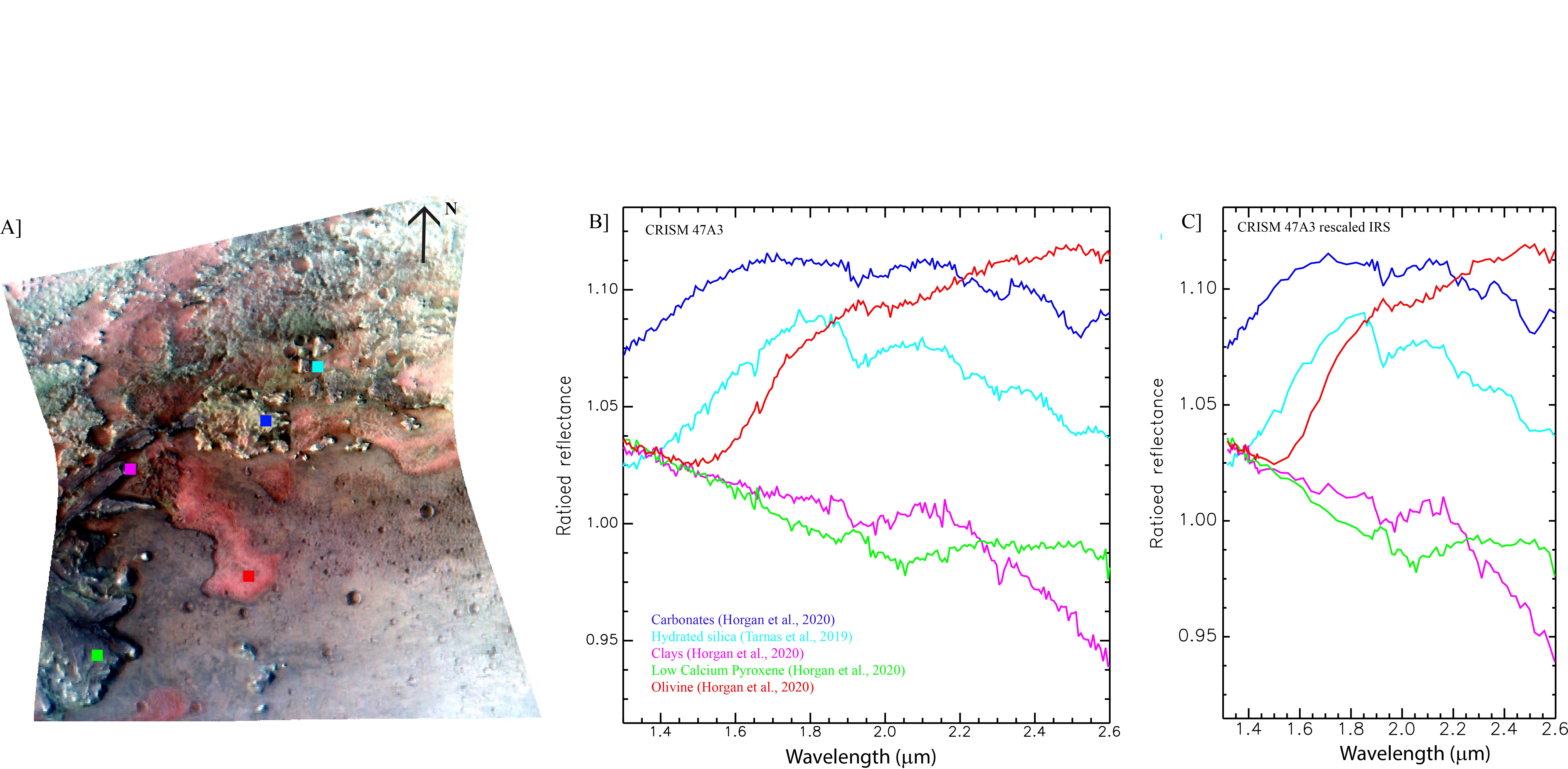}
	        \caption{Examples of spectra representative of the mineralogical diversiry encountered in Jezero crater extracted from the CRISM data cube FRT000047A3\_07\_IF166L\_TRR3. The data cube has been processed using CRISM Analysis Toolkit \citep[CAT,][]{Murchie2007} via MarsSI plateform \citep{Quantin2018}. The processing includes calibration, atmospheric contribution removal, denoising, column median ratioing and map-projection.  A] RGB composite (R: 2.529~$\mu$m; G: 1.5066~$\mu$m;B: 1.08~$mu$m); the color squares indicate the location of the spectra displayed in B and C. B] CRISM ratioed spectra (average of $3\times3$ pixels window) of identified spectral signatures in literature. C] CRISM ratioed spectra resampled to Supercam infrared spectrometer Table\#4 (see Tab~\ref{NaccTable}.}
	        \label{FigMineralDiversity}
	    \end{figure*}

	    In summary, we expect to investigate a large diversity of igneous minerals or alteration minerals once Mars2020 has landed within Jezero crater. Figure~\ref{FigMineralDiversity} displays this mineral diversity, highlighting some CRISM spectra convolved and sampled at the Supercam infrared spectrometer spectral resolution and spectel wavelengths.
	    
    \subsection{Mars atmosphere in the near infrared}
    \label{SecMarsAtmosphere}
    
        The gases that shape the transmission of the Martian atmosphere in the near infrared are mainly CO$_2$, water and CO. Since CO$_2$ (i.e.\ surfarce pressure) will not be measured more accurately by the SuperCam infrared spectrometer than by dedicated instruments on board Perseverance, we concentrate here on water and CO.

        \begin{figure*}
            \centering
            \includegraphics[width=0.75\linewidth, clip=true, trim=0 70mm 0 65mm]{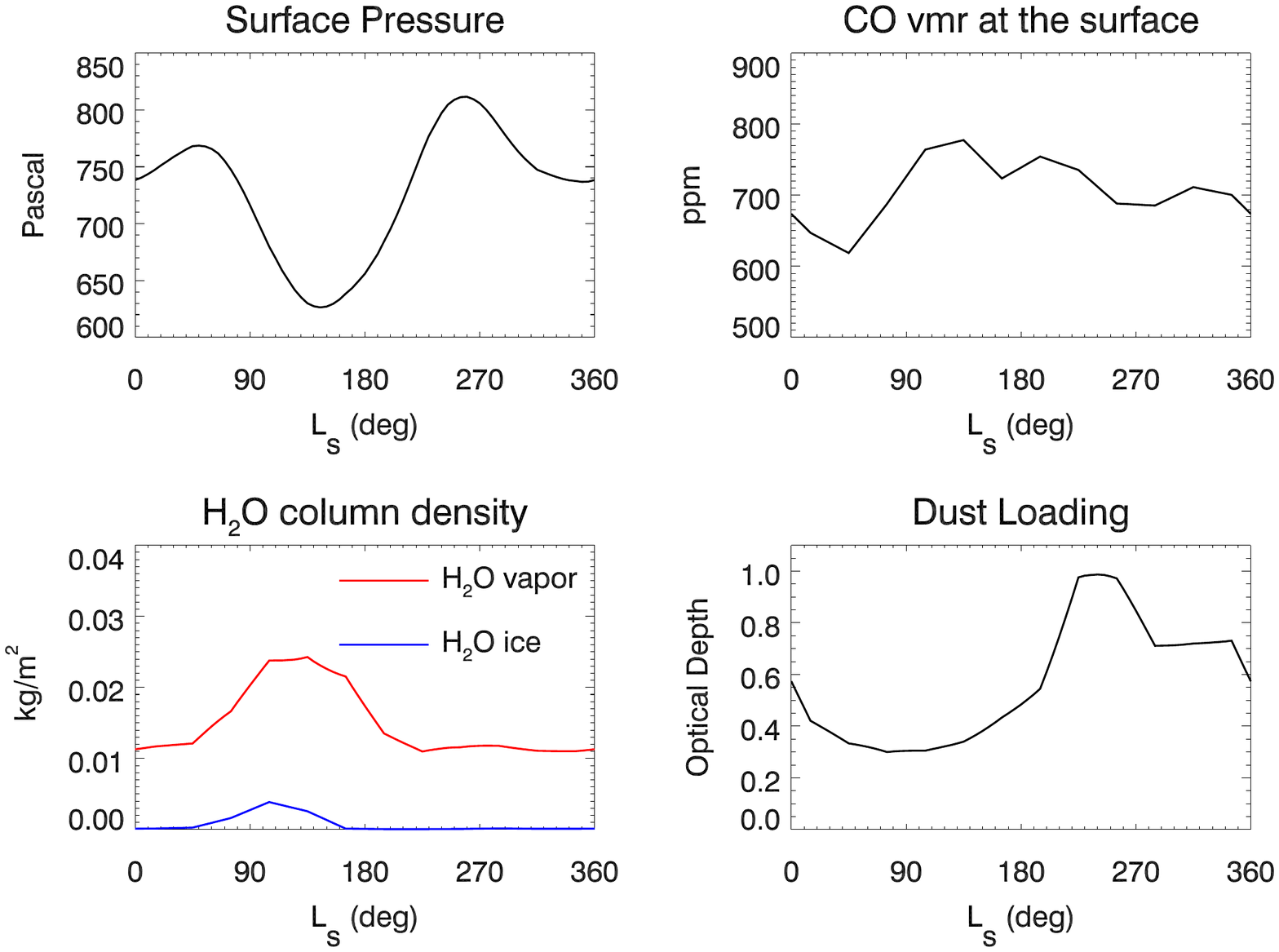}
            \caption{Jezero crater climatology of Surface Pressure (upper left panel), CO abundance 1 meter above the surface (upper right panel), H$_2$O vapour (red line) and H$_2$O ice aerosol (blue line) column densities (lower left panel), and dust column opacity (lower right panel) extracted from the Mars Climate Database \citep[http://www-mars.lmd.jussieu.fr/mcd\_python/]{Forget1999,Millour2019} for the average solar scenario.}
            \label{FigClimatologie}
        \end{figure*}
        
        The water cycle (lower left panel of Fig.~\ref{FigClimatologie}) was first fully described using MAWD/Viking data and then surveyed with TES/MGS observations, followed by those of Mars Express with the contributions of the SPICAM, OMEGA, and PFS instruments (see the review by \citet{montmessin2017} and references therein). More recently, CRISM/MRO and NOMAD/TGO, using near-infrared spectroscopy, also contributed to the regular and continuous monitoring of the water cycle on Mars \citep{Smith2018,Vandaele2019,Aoki2019}. All instruments concur to describe a water cycle in which the North Polar Ice Cap is the main source of water by sublimation during the northern summer, and in which the asymmetry of the seasons between the two hemispheres confines water to the Northern Hemisphere, in agreement with the models \citep{montmessin2017}. In the northern polar region, the summertime water column density is on average about 50~pr.-$\mu$m with local peaks at 90~pr.-$\mu$m. Transported to the northern tropical regions, water is diluted to 15~pr.-$\mu$m. In the southern hemisphere, the summertime peak reaches about 30~pr.-$\mu$m, while the tropical water remains low as water is efficiently transported to the northern hemisphere. This annual cycle is extremely regular from year to year, with only global dust storms affecting its regularity \citep{Smith2004,Fouchet2011,Fedorova2018,Vandaele2019}. The role of atmosphere-surface exchanges in the water cycle has been heavily debated (see \citet{montmessin2017} and references therein). In this regard, the comparison of the REMS/MSL humidity measurements in the vicinity of the Curiosity rover \citep{Harri2014b} and the ChemCam/MSL column density measurements \citep{McConnochie2018} supported the possibility of substantial exchanges \citep{Savijarvi2016,Savijarvi2019}.
    
        Deciphering the cycle of carbon monoxide (upper right panel of Fig.~\ref{FigClimatologie}) was more difficult than that of water vapour. Early tries by the Mars Express instruments had not grasped the full seasonal behaviour of CO \citep{Encrenaz2006,Billebaud2009}, before CRISM observations revealed an annually and planetary averaged carbon monoxide mixing ratio of 700 ppm, with strong seasonal and latitudinal variations, dropping at summertime lows of 200 ppm in the southern hemisphere and 400 ppm in the northern hemisphere \citep{Smith2009}. These variations are caused by the dilution of CO as a noncondensable species when carbon dioxide sublimates from the polar caps during summers.
    
        Besides gases, clouds and dust play also a very important role in shaping Mars near-infrared spectrum, and as a consequence, in the Martian atmospheric radiative budget.
    
        Dust was identified as the leading positive radiative forcing agent in the 1970s, and has since been monitored and studied very regularly for this reason. As reviewed in \citet{kahre2017}, TES/MGS, THEMIS/MRO and MCS/MRO have followed the dust seasonal cycle very efficiently and allowed several models to establish a long-based climatology of dust optical depth now incorporated as the nominal dust scenario in the Mars Climate Database (see lower right panel of Fig.~\ref{FigClimatologie}). Apart of Martian years affected by global encircling dust storms, the dust seasonal cycle is quite stable from year to year. The dust opacity remains low during the northern spring and summer, then gently rises in the southern hemisphere at $L_s=180^{\circ}$, before largely increasing between $L_s= 220^{\circ}$ and $260^{\circ}$, abating between $L_s=250^{\circ}$ and $300^{\circ}$ except in the southern polar region, and then surging again between  $L_s=250^{\circ}$ and vernal equinox. Multiple observations from surface-based rover have surveyed the same global pattern, but also revealed local scale meteorology (\citet{kahre2017} and references therein). For example in Gale crater, Curiosity measurements have shown that the atmosphere within the crater efficiently mixes with the free atmosphere only around $L_s=270^{\circ}$--$290^{\circ}$ \citep{Moore2016}.
    
        The knowledge of the dust photometric parameters (single scattering albedo, phase function) is also of prime importance to precisely model the dust radiative forcing. Following the early work of \citet{Ockert-Bell1997}, \citet{Wolff2009} using CRISM/MRO, and \citet{Maattanen2009} using OMEGA/MEX retrieved the single scattering albedo of the martian dust across the full visible and near-infrared domains.
        The dust phase function was also accurately measured from TES/MGS observations \citep{Wolff2003}, but the forward scattering properties have never been studied from the surface in the near-infrared. Orbital and surface-based measurements both show that the mean particle radius increases with the dust opacity \citep{Clancy2003,Wolff2003,Smith2016,Lemmon2019} affecting the dust scattering properties.
    
        The first identification of water ice in Martian clouds was performed in the thermal infrared by Mariner 9, before the TES, THEMIS and MCS instruments allowed for a complete survey of the water ice clouds seasonal cycle (see \citet{clancy2017} and references therein). The water cycle is marked by the aphelion cloud belt during the northern hemisphere summer and polar hoods during polar winters (Lower right panel of Fig.~\ref{FigClimatologie}). In this context, \citet{Madeleine2012} were the first to monitor the 1.5, 2.0, and 3.1-$\mu$m absorption bands of water ice clouds using OMEGA/MEX. From the Martian surface, the MER rovers and Curiosity, using their navigation cameras also followed the seasonal, and local, behaviour of the martian water ice clouds \citep{Lemmon2015,Moore2016,Kloos2018}. Martian clouds play an important role in the radiative budget of the martian atmosphere, heating the atmosphere in daylight and cooling the atmosphere during nighttime \citep{Madeleine2012}, a role that is supposed to have been more important for ancient Martian climates. To model this radiative effect, it is important to determine the water ice cloud phase function, which has been done mostly in the UV, visible and thermal infrared domains \citep{clancy2017,Cooper2020} but not performed in the near-infrared.
 
\section{Scientific objectives}
\label{Scientificobjectives}

    The SuperCam infrared spectrometer will contribute to the four overarching science objectives of the Mars 2020 mission and its Perseverance rover \citep{Mustard2013,Farley2020}:
    \begin{enumerate}
        \item Mars2020 should develop a global understanding of the geology of its landing site;
        \item Mars2020 should identify ancient habitable environments, estimate the capability of rocks to preserve biosignatures, and identify potential biosignatures;
        \item Mars2020 should collect and document several rock samples for possible Earth return;
        \item Mars2020 should enable future Mars exploration especially by humans, and by demonstrating new technologies.
    \end{enumerate}
    As detailed in \citet{Maurice2020}, these mission's objectives have been declined in eight different scientific goals for the SuperCam instrument:
    \begin{itemize}
        \item Goal 1: Rock Identification
        \item Goal 2: Sedimentary Stratigraphy and Facies / Hydrothermal Characterization.
        \item Goal 3: Organics and Biosignatures
        \item Goal 4: Volatiles (Hydration and Halogens)
        \item Goal 5: Context Morphology and Texture
        \item Goal 6: Coatings and Varnishes
        \item Goal 7: Regolith Characterization
        \item Goal 8: Atmospheric Characterization
    \end{itemize}
    
    The SuperCam infrared spectrometer will be a key instrument to achieve Mission Goals 1 and 7, through its capacity to identify mineral constituents of the rocks. The detection capacities of the IRS are presented in Tab.~\ref{TabDetectionLimit} and detailed in Sec.~\ref{SecMineralIdentification} for various families of minerals and for some organic species. As detailed in Sec.~\ref{SecAtmosIRS}, the IRS will also contribute to Goal 8 - Atmospheric Characterization, and more marginally to Goal 3 - Organics and Biosignatures (Sec.~\ref{SecOrganics}).
    
    \begin{table}
        \caption{Table of key-minerals and organics targeted by the SuperCam Spectrometer in their specific absorption bands .}\label{TabDetectionLimit}
        \centering
        \begin{tabular}{lll}
            Mineral family & Examples &  Signature Wavelength (µm) \\
            \hline
            Carbonates & Calcite & 2.35, 2.55\\
            & Aragonite & 2.35, 2.55 \\
            & Magnesite & 2.30, 2.50 \\
            \hline
            Chain silicates & Pyroxenes & 2.0 \\
            & Olivines & 1.0 \\
            \hline
            Sheet silicates & Serpentines & 1.4, 2.0, 2.1, 2.3, 2.5\\
            & Smectite & 1.42, 1.92, 2.2/2.3\\
            & Talc & 1.37, 2.30\\
            & Kaolinite & 1.40, 2.22\\
            & Zeolithe & 1.4, 1.9, 2.5\\
            \hline
            Sulfates & Gypsum & 1.45, 1.94, 2.20\\
            \hline
            Organics & Bitumens & 1.7, 2.15, 2.3--2.5\\
            & Coals & 1.7, 2.3, 2.5\\
            & CN Bonds & 2.3\\
            & Alcane & 2.3\\
            & Coronene & 1.7, 2.1-2.6\\
            \hline
    \end{tabular}
    \end{table}

    \subsection{Mineral identification}
    \label{SecMineralIdentification}
    
        \subsubsection{Carbonates}
        \label{SecCarbonates}
        
            The carbonate minerals can be identified primarily through the overtones and combinations of their C-O stretching and bending modes ($3\nu_3$ at 2.3~$\mu$m and $\nu_1 + 2\nu_3$ at 2.5~$\mu$m). The exact wavelengths of the mode centers can help to identify their major cations. For example, Mg-rich carbonates exhibit minima at relatively shorter wavelengths than Ca-rich and Fe-rich carbonates. In addition, carbonates can exhibit a hydration band at 1.9--2.0~$\mu$m of various forms and centers depending on the type of hydration (OH/H$_2$O in hydrocarbonate, adsorbed water, fluid inclusion).
            
            In Jezero and Nilli Fossae, carbonates are mostly present in association with olivine in the olivine/Mg-carbonate rich terrains that also exhibit hydration bands. This coexistence of the infrared olivine and carbonate signatures is spotted in Jezero's watershed, specifically in the mottled terrains, but also in the crater rim, the delta and light-toned floor within the crater \citep{Ehlmann2008,Ehlmann2009,Brown2010,Goudge2015,Salvatore2018,Horgan2020}. The exploration of these different units and terrains will reveal if their olivine/Mg-carbonate phases have the same source pre-existing Jezero formation, with minerals latter transported by fluvial or aeolian processes to the delta, and crater rim and floor, or if these different terrains present separate and different episodes of alteration, each one possibly originating from a different mechanism.

            Indeed, infrared spectral signature will contribute to inform us about the formation mechanism and the environmental and geochemical conditions of their formation. As outlined by \citet{Ehlmann2011} different minerals should have formed from the aqueous alteration of igneous rock in a closed, subsurface, environment than in an open, surface, environment. In a closed environments, at low temperatures, basalts are preferentially altered to iron oxides and smectites, while higher temperatures lead to chlorite, serpentine, or amphiboles \citep{Franzson2008}. In open environments, iron and magnesium are transported away and the alteration leads to the formation of Al-rich phyllosilicates \citep{Hurowitz2007}. The relative enrichment in Al over Fe/Mg depends on the magnitude of the alteration. The mineral assemblage of carbonate will also give clues on the weathering process (serpentinization or carbonation), and the temperature and pressure conditions of alteration \citep{Kelemen2020}. Sampling and eventual return of this olivine/carbonate unit will constrain the possibility of microbial life in this sediment \citep{Brown2020} through the use of carbon isotopic ratios. Geochronological determinations of carbonate ages will help understand whether carbonates on Mars could be the dominant reservoir of the early thick, CO$_2$-rich Martian atmosphere \citep{Edwards2015}.
            
            Carbonates are also widely present in the marginal carbonate unit to the north and south of the delta, in a form that geographically conforms with the former lake shoreline. This marginal carbonate unit exhibits a wider diversity of spectroscopic signature than does the olivine/carbonate unit \citep{Horgan2020}, mixed with various degree of hydration and the presence of olivine signature. The SuperCam infrared spectrometer will explore the spectral diversity of the carbonate signature in this unit, and correlate it with grain size and shapes, mineralogical assemblage, facies and laminations observed in images to help disentangle various origins, such as authigenic lacustrine precipitation, local deposition of materials from the watershed, or in place alteration of an olivine-rich unit \citep{Horgan2020}. If an authigenic lacustrine deposition is favored, possibly biomediated, the exploration of the marginal carbonate unit will arise as a major objective for the Perseverance rover, due to the high biosignature preservation potential of lacustrine carbonates. 
 
        \subsubsection{Chain silicates}
        
            Ortho- and chain-silicates present large absorption bands around 1.0 and 2.0~$\mu$m, due to Fe$^{2+}$ crystal field transitions \citep{Adams1974,King1987,Burns1993}. The presence or absence of the 2.0-$\mu$m band allow mineralogists to discriminate pyroxene from olivine, while its position strongly depends on the nature of the pyroxene, preferentially at 1.9~$\mu$m for low-calcium pyroxene (LCP), and 2.2-2.3~$\mu$m for high-calcium pyroxene \citep{Klima2007,Klima2011}. The SuperCam infrared spectrometer is not well suited to detect olivine because of its cut-off wavelength at 1.3~$\mu$m, but still can provide clues of olivine signature, especially in conjunction with the SuperCam Body Unit spectrometers (covering the $\sim$400--850~nm region), as the olivine and some HCP 1.0-$\mu$m-band is broad enough to be partially sampled in its visible and infrared wings.
            
            Generally, the remote identification of igneous minerals by the SuperCam infrared spectrometer will be decisive to guide the Perseverance rover towards rocks that might have preserved the history of Mars accretion, differentiation, and early crustal processes, and which will be prime targets for returning samples of geochronological interest. Their presence at various places, in the crater floor, crater rim and Jezero watershed will give the opportunity to explore and sample different epochs and different types of volcanism.
            
            Specifically, low-calcium pyroxene is present in the Jezero crater rim and outside Jezero \citep{Goudge2015,Horgan2020}. These are  thought to be representative of the deep crust or even the upper mantle of early Mars uplifted or excavated by the Jezero and Isidis impacts. Samples from this unit analysed in Earth laboratories could constrain the absolute epoch of the crust formation, the chemical and isotopic composition of reservoir from which the early Martian crust formed \citep{Mustard2009,Ody2013}. They will help to assess whether the Martian crust is heterogeneous, and what its relative composition in volatiles is, compared to other terrestrial bodies. These samples could also have recorded the history of magnetism on Mars.
            
            Low-calcium and high-calcium pyroxenes have also been identified in the Jezero floor, in a dark-tone unit that overlies carbonate‐bearing deposits; \citet{Goudge2015,Horgan2020} interpreted this dark tone unit as a lava flow. Sampling of this unit will be important to understand the timing of the Jezero history and to calibrate the crater chronology at relatively recent epochs. Indeed, different studies based on crater counting yielded widely different results: \citet{Schon2012} favored an early Amazonian age of 1.4~Ga, \citet{Shahrzad2019} a late Hesperian age of $2.6\pm0.5$~Ga, while \citet{Goudge2012} found an early Hesperian age of $3.45^{+0.12}_{-0.67}$~Ga. Exploration of the LCP and HCP units in the Jezero floor will also be important to understand how the Martian volcanism has evolved from LCP to HCP in the amazonian period \citep{Mustard2005,Quantin2012}. 
            
            The olivine-carbonate unit present outside and inside the Jezero crater floor will also be a prime target to date igneous processes in various Martian environments \citep{Hoefen2003,Brown2010,Goudge2015,Horgan2020}. One question is whether the olivine-carbonate rocks present in the Jezero crater floor, the delta, the Jezero crater rim and outside Jezero are different expressions of the same unit and whether this unit is due to ash flows, lava flows or impact melts \citep{Mustard2007,Kremer2019,Brown2010}. We expect to find the oldest expression of this olivine-carbonate unit outside Jezero, which was dated by crater counting to trace back to $3.82\pm{0.07}$~Ga by \citet{Mandon2020}. The distinct composition of olivine-carbonate will give clues on the evolution of the martian interior composition as has been performed by \citet{Edwards2017} using ChemCam who found the magma in Gusev and Gale craters similar, and different from Shergotite and Tharsis magma.
        
        \subsubsection{Sheet silicates}
        \label{SecSheetSilica}
        
            Sheet silicates (smectite, serpentine, talc, kaolinite) can be detected through the first harmonic of the hydroxyl fundamental vibrational mode at 1.4~$\mu$m, and through its combination with the transverse vibrational modes of, for example Al-OH at 2.2~$\mu$m, or Fe/Mg-OH at 2.3~$\mu$m \citep{Clark1990,Bishop2002,Frost2002}.
            
            In the Nili Fossae region, Mg/Fe-smectite rich material has been identified at several places, including within the Jezero crater rim and crater floor, and within the delta \citep{Mustard2009,Ehlmann2008,Ehlmann2009,Goudge2015,Horgan2020}. In-situ exploration of the Mg/Fe-smectite rich material at these different places will seek to determine whether this mineral is either igneous or sedimentary in origin\citep{Scheller2020}, and if the different occurrences represent detrital units related to an early-Noachian implacement or different authogenic/diagenic expressions. If igneous, the Mg/Fe unit outside Jezero and in the uplifted basement within the crater rim will give access to crustal processes in early Mars (early Noachian, possibly pre-Noachian). If sedimentary, this unit will provide clues on the geochemical and climatological environments on early Mars. The SuperCam infrared spectrometer will allow us to determine the extent of the alteration, and will guide the Perseverance Rover to collect samples that will be instrumental in determining the nature of the aqueous alteration in early Noachian times, the precise age of the alteration and the isotopic conditions at that epoch.
            
            Another mineral that will be instrumental in the determination of the alteration conditions on early Mars is kaolinite. Kaolinite-bearing blocks have been spotted within the Megabreccia of the Jezero crater rim \citep{Cuadros2013}. This Al-rich phyllosilicate is extremely useful to disentangle the conditions of the alteration on early Mars, i.e.\ surface versus hydrothermal alteration, as the ratio of the 2.16~$\mu$m to 2.20~$\mu$m band depth gives insight on its cristallinity \citep{Kelemen2020}. A returned sample of kaolinite-bearing blocks could provide strong constraints on the temperature and pressure conditions of the alteration that took place on early Mars.
            
        \subsubsection{Sulfates}
        
            Sulfates, either monohydrated or polyhydrated, will be sought using the harmonics and combinations of OH- or H$_2$O bending and stretching modes at 1.4~$\mu$m \& 1.9~$\mu$m, and through the $3\nu_3$ harmonic of the SO$_4^{2-}$ radical at 2.4~$\mu$m \citep{Cloutis2006}.

            Sulfate minerals have been spotted in the extended Nili Fossae region \citep{Ehlmann2009}, in a layered sedimentary unit overlying the regional mafic capping unit, but outside of the Jezero watershed and Jezero crater. Hence, these minerals have not yet emerged as a major objective for the exploration of Jezero crater. However, the experience of in-situ Mars surface exploration by rovers reveals that small scale calcium-sulfate veins are as ubiquitous on the Martian surface even though they are undetectable from orbit \citep[see also Sec.~\ref{SecScale}]{Squyres2012,Arvidson2014, Grotzinger2014}. These veins are mostly composed of gypsum and bassanite, the latter originating from the dehydration of former \citep{Nachon2014,Rapin2016}. Chemcam observations analyzed by \citet{LHaridon2018} also showed that these calcium-sulfate veins sometimes preserve signatures of the evolution of acidity and redox conditions of martian fluids.
            
            Calcium-sulfate veins have been proposed to have a high biological interest. For example, \citet{Benison2014} showed that acidophilic organisms can develop and be trapped within fluid inclusions in Mars-analog gypsum. \citet{Gasda2017} also highlighted that boron detected by ChemCam in Ca-sulfate veins is thought to play a key role in the emergence of life. The SuperCam infrared spectrometer will provide the Perseverance rover with the capacity to determine the hydration and the nature of sulfate minerals and to decide on the most suitable ca-sulfate for sample and return for investigation on Earth.
            
        \subsubsection{Organics}
        \label{SecOrganics}
        
            Finally, the SuperCam infrared spectrometer will have the possibility to detect organics through overtones of the CH, C$=$C or CN vibrational modes at 1.7, 2.15, 2.3, and 2.5~$\mu$m, but only if they are present in very large quantities. The organics detection could be made more difficult by the fact that their preservation since $\sim3.5$~Ga or more may have involved chemical evolution like sulfurization as highlighted by MSL \citep{Eigenbrode2018}.
            
    \subsection{The change in scale from orbit to ground exploration: the fine (grain) scale}
    \label{SecScale}
    
        The importance of change of scales has been demonstrated nicely with the Curiosity rover in Gale Crater. The most prominent example of this change of scale is the observation of phyllosilicates by the CheMin instrument in drill holes along most of the Curiosity traverse, starting with the first drill holes at Yellowknife Bay \cite{Grotzinger2014}, while orbital spectra did not detect phyllosilicates except at the clay bearing unit on the side of Aeolis Mons. Thus, in situ exploration revealed that the presence of phyllosilicates in Mars sediments is likely significantly higher throughout Mars than observed from orbit. The large set of veins that has been observed with the imaging camera Mastcam \citep{Kronyak2019} and ChemCam RMI \citep{Nachon2014} constitutes another example of change of scale. It appeared that those veins were composed mainly of Ca-sulfates \citep{Nachon2014}, \citep{Nachon2017}. Those Ca-sulfates were not detected from orbit. One explanation could rely on the fact that those veins were partially obscured by the Martian dust, and therefore less detectable from orbit with the operating IR spectrometers, which are very sensitive to the iron contained in the dust. A final example of change of scale is illustrated by the huge signal of hematite detected from orbit by CRISM on the so called “Vera Rubin Ridge” (VRR) \citep{Fraeman2020}. Although the mean iron content turned to be not particularly more elevated than elsewhere around Mount Sharp \citep{Fraeman2020}, it appeared that hematite was detected by the ChemCam passive spectroscopy in small patches and concretions \citep{David2020,LHaridon2020}. These two examples illustrate the fact that the dust coverage plays an important role in obscuring the spectral signatures and that it is crucial to have access to the small scale to be able to locate precisely the textures that potentially bear the signatures observed from orbit. The possibility of SuperCam to remove the dust using laser shots before the acquisition of passive spectra with IRS will be of significant importance to mitigate the dust masking effect.   

        In Jezero crater, the question of the origin of the marginal carbonates detected from orbit \citep{Horgan2020} is critical and the IRS will provide clear distinction between the various types of carbonates in particular if they are Mg or Ca-dominated, or even if they are enriched in Fe thanks to the relative positions of the 2.3-$\mu$m and 2.5-$\mu$m bands (see Sec.~\ref{SecCarbonates}). These three carbonate-bearing units of the crater’s margin are light-toned and fractured, but do exhibit some differences in smaller scale surface textures. These textural differences suggest that these terrains experienced different origin, alteration, or erosional processes. However, the nature of the difference is unclear from texture alone and the small footprint of the IRS instrument will tell how these carbonates were emplaced. For example, \citet{Caldirak2018} showed that magnesite is diffuse in sands, and that it can form large, pure, dense, and hard nodules. The IRS will easily discriminate between these possibilities. These carbonates are often associated with hydrated phases that vary independently from them. These phases could be clays or hydrated silica. The IRS instrument will be able to distinguish between Fe/Mg/Al clays and hydrated silica based on the locations of the various metal-OH transitions. It has been assumed that some serpentinization processes could produce these phases. The IRS instrument will be able to discriminate between serpentine, talc, and opal, all phases being produced at various stages of the serpentinization process and, consequently give information on the thermodynamic conditions leading to their formation. The footprint of the IRS will be characterizing the setting of these phases (veins, nodules, interlayered strata, …) and distinguishing between various crystal growth in direct chemical precipitation from surface water.
        
        Other hydrated phases like kaolinite --- possibly present in the mega-breccia of the crater rim --- will be easily identified (see Sec.~\ref{SecSheetSilica}). It will be possible to characterize alteration profiles of peridotites, possibly sepentinized with the presence of magnesite, to Fe-Mg-Al smectite and then kaolinite (\citep{Gaudin2011},\citep{Brown2020}). However, kaolinite also occurs in secondary deposits where it, or its parent minerals, have been transported under suitable non-alkaline conditions and deposited in deltaic, lagoonal, or other non-marine environments like the Jezero delta. In any case the IRS instrument will be able to distinguish between the various alteration processes and their thermodynamical properties by clearly identifying the mineral alteration assemblages. An in situ investigation of the Mg/Fe-phyllosilicate mineralogy and the nature of the hydration event is therefore a critical task in understanding the astrobiological potential of the carbonate and phyllosilicate deposits.

    \subsection{Monitoring the Martian atmosphere from the surface}
    \label{SecAtmosIRS}
    
        The SuperCam infrared spectrometer will participate in the efforts of the Mars2020 mission to enable future Mars human exploration by filling gaps in some strategic knowledge on the behaviour of the Martian atmosphere, and on the Martian climate and meteorology. The findings made by the SuperCam infrared spectrometer will also serve the general quest of Perseverance to understand the past Martian habitability by providing key-parameters for paleoclimate numerical simulations. In particular, the SuperCam infrared spectrometer will provide crucial paramaters for numerical modelling of the current and past Martian climates, and will also monitor the Jezero local weather to help in improving operational meteorological forecasts.
        
        Most importantly, the SuperCam infrared spectrometer will monitor the sky brightness at different elevations and azimuths, hence phase angles in the full 1.3--2.6~$\mu$m range. From these observations, it will be possible to retrieve the dust and water ice (when clouds are present) phase functions. These phase functions will be determined for the first time from the ground in the near-infrared domain. This measurement will be very important for radiative models of the Martian Atmosphere.
        
        Water ice clouds also currently contribute to this radiative budget, but heir role might have been more important in the past \citep{Wordsworth2013,Turbet2017}. Combining SuperCam Body Unit spectral coverage together with the IRS spectral coverage, from the UV domain up to 2.6~$\mu$m will be instrumental in investigating how the radiative properties evolve with the dust particle size, in determining the dust and ice size distributions. These SuperCam observations will be compared to, and analyzed with MastCam-Z and NavCam observations, which will provide a broader field-of-view (FOV) for sky observation than the SuperCam infrared spectrometer, acting as point spectrometer, can perform.
        
        The SuperCam infrared spectrometer will measure the column-integrated H$_2$O and CO abundances, as well as the column-integrated dust opacity. Comparison with local measurements performed by other instruments will reveal possible local meteorological processes. Using the 1.9 and 2.6-$\mu$m water vapour band, SuperCam's infrared spectrometer will complement the water vapor measurements that can be performed by the SuperCam Body Unit spectrometers, resulting in a improvement of sensitivity and precision compared to ChemCam measurements \citep{McConnochie2018}. For standard atmospheric conditions, SuperCam infrared spectrometer measurements will achieve a detection limit of 3~pr.-$\mu$m and a precision of $\pm0.5$~pr.-$\mu$m. The comparison between the local measurements made by MEDA in the vicinity of the rover and the column-integrated measurements performed by SuperCam will reveal the local exchanges of water between the surface and the atmosphere \citep{McConnochie2018}. As with mineralogy, the dustiness of Jezero crater differs from that of the other environments explored by rovers. Hence combining the results in different environments will test the hypothesis that water adsorption by the Martian regolith plays an important role in the planetary water cycle \citep{Bottger2005}, or will find that it plays only a local role. The SuperCam precision on the CO column abundance will be limited to 200 ppm for a standard spectrum. Several successive acquisitions will be required to achieve a better precision. It will be especially interesting to monitor any CO variability associated with deviations of O$_2$ from the expected seasonal cycle \citep{Trainer2019}, as O$_2$ measurements with SuperCam’s Body Unit spectrometers are anticipated in the same manner as with ChemCam \citep{McConnochie2018}.
        
        The IRS will also measure the dust column-integrated opacity by scanning the sky brightness. From a photometric point of view, a precision of $\Delta\tau=\pm0.01$ can be achieved on this measurement, but other parameters, such as the vertical and horizontal heterogeneity of the dust distribution, will affect the data analysis and prevent this $\Delta\tau=\pm0.01$ precision from being achieved. Comparison between local and column-integrated measurements made by the Curiosity rover revealed how the atmosphere in Gale crater just occasionally mixes with the free Martian atmosphere.
        Along with MEDA, and MastCam-Z, it will test if the atmosphere within Jezero crater mixes with the free atmosphere only at some specific $L_s$ as it occurs in Gale crater. Since the two craters differ in rim altitude and diameter, comparison between the meteorology in these two different environments will further test the ability of mesospheric numerical models to predict precisely the local meteorology on Mars \citep{Spiga2018}. This ability of mesospheric models in predicting the weather at various Martian locations will be crucial in the future human exploration, for example to select landing sites where the safety of the facilities and personnel will be guaranteed. 
        
    \subsection{The synergy with the SuperCam instrument suite and the Mars2020 payload}

        \subsubsection{Synergy within the SuperCam instrument suite}
        
            SuperCam will provide nested and co-aligned investigations of the same target with four different spectroscopic and imaging techniques that will altogether extensively and thoroughly characterize the mineralogy and the geology of the landing site \citep{Wiens2020,Maurice2020}.
            
            \paragraph{Raman and time-resolved fluorescence spectroscopy}
            Infrared and Raman are spectroscopies that interrogate the vibrational structure of molecules, materials or minerals. These techniques are often combined in laboratory studies as they give access to complementary information. However, they are based on two distinct physical processes: absorption for infrared light versus inelastic scattering for Raman. In the case of SuperCam, both techniques will have a comparable FOV, 0.74~mrad for Raman and 1.15~mrad for IR, so analytical footprints from about 1.5~mm at 2~m to about 5.2~mm at 7~m, which is significantly larger than the spot size for LIBS on the target. On many fine- to medium-grained targets, these techniques will probe a mixture of possibly various minerals. Both techniques are highly sensitive to internal vibrations of molecular groups (e.g. C-O, O-H…) in minerals but Raman can also characterize external lattice vibrations giving access to the crystal structure. This gives to Raman the possibility to distinguish mineral polymorphs, e.g. Ca carbonates: calcite versus aragonite. In addition, the two techniques do not have the same efficiency depending on the minerals: some minerals will strongly absorb in the near-infrared light whereas others will instead have an excellent Raman efficiency. Consequently, some minerals will be preferably characterized by infrared like phyllosilicates whereas some will be more appropriate for Raman spectroscopy like phosphates, sulfates or some framework silicates (e.g. quartz, plagioclases…). Interestingly, carbonates and some silicates (e.g. pyroxenes, olivine) will have a specific spectral signature with both techniques providing a double assessment and also complementary information for in-depth mineralogical characterization. In addition, both infrared and Raman can detect hydrous ices as well as organics (from molecules to kerogen or graphitic materials) and may provide important clues for their identification.
            
            \paragraph{LIBS Spectroscopy}
            
                The LIBS and VISIR techniques have unique and different capacities. LIBS yields unique quantitative elemental compositions for all the major rock-forming oxides, and measures most of the minor metallic and non-metallic elements (H, N, P, S, and less efficiently C and O) \citep{Maurice2020}. On its side, VISIR yields unique information on the mineralogical composition of the rocks (see Sec.~\ref{SecMineralIdentification}). Hence, the combinations of elemental chemistry and VISIR spectra will be very complementary. On MSL, having only elemental chemistry at stand-off distances leaves many questions unanswered as all mineralogy must be inferred from chemistry. This forces the team to rely mostly on the chemical index of alteration \citep[CIA,][]{Nesbitt1982} to only very partially understand authigenic alteration. Having both VISIR and LIBS will allow the level of alteration to be understood firsthand, from the mineral spectra. In this way, the chemical index of alteration can be double-checked on Mars, which has not been directly possible except in a few instances.
                
                In addition, VISIR will be the unique rock-interogation technique farther than 10-m away from the rover, while LIBS will be the only effective technique whenever grain size, morphology, or shadows may render VISIR spectra more difficult to interpret. LIBS chemistry can also highlight minor mineral phases or phases that are not strongly diagnostic in VISIR or Raman spectra. 
                
                The footprint of LIBS is substantially smaller than VISIR and Raman spectra, so  interpretation of the combined data sets will be open to attributing inferred differences to different footprint sizes. However, the large number of data points expected from SuperCam should help elucidate any perceived inconsistencies.
                
                Finally, the LIBS analysis will clear the dust of the targets before they are analysed with VISIR, therefore providing spectra with better discriminating capabilities.
                
            \paragraph{Remote Micro-Imaging}
            
                The RMI will provide context images for IR (as well as LIBS and Raman) targets. Compared to MSL/Chemcan \citep{LeMouelic2015}, the possibility to obtain color RMI images will enable comparison with IR mineralogical spectra. With an IFOV better than 80~$\mu$rad/pixel, RMI resolves individual sand-size grains ($\sim$25 to 250~$\mu$m/pixel at 1.2 to 12~m distance, respectively) allowing detailed texture investigations. RMI will inform us on grain size, which will be valuable for IR mineralogy modelling with radiative transfer. The RMI images will also allow us to correlate textures and facies with mineralogy, which will inform whether minerals are preferentially authigenic or detrital, igneous or sedimentary. It will document mineral arrangements at the submillimeter level that will be important to identify potential biosignatures.  Additionally, RMI images will support passive VIS and IR spectra as one of the primary means of long-range mineral identification of key targets for broad geological interests. Hence these combination of remote sensing techniques will be a key asset for Mars2020 strategic and tactical planning.

        \subsubsection{Synergy with the Perseverance payload}
        
            The SuperCam infrared spectrometer will work very closely with the three other Mars2020 instruments devoted to investigating the landing site mineralogy: Mastcam-Z, SHERLOC and PIXL.
            
            Like IRS, Mastcam-Z will achieve mineralogical detection at the millimeter scale in the vicinity of the rover and centimeter scale at km distances with an IFOV ranging between 67~$\mu$rad to $0.283$~mrad. Mastcam-Z will be especially complementary to the IRS and the SuperCam BU spectrometers, by covering the 1-$\mu$m absorption band of Fe$^{2+}$-bearing and Fe$^{3+}$-bearing minerals. Hence, it will partly compensate the IRS lack of sensitivity to the 1~$\mu$m region for minerals such as low-calcium and high-calcium pyroxenes, olivine, and Fe-rich phyllosilicates like nontronite. With its larger FOV and multi-spectral imaging capacity, Mastcam-Z will also extend the mineralogy locally assessed by the IRS to the outcrop scale, following layering and stratification.
            
            At the sub-millimeter scale, the mineralogical composition of rocks will be investigated by SHERLOC and PIXL. SHERLOC, a UV-excited Raman and fluorescence spectroscopy instrument, is designed to be much more sensitive to organics than the IRS. For minerals, SHERLOC will extend the detection and mapping of OH-rich minerals, sulfates and carbonates down to the grain size. However, its cutoff at 800~cm$^{-1}$ for the Raman shift, will make it challenging to compensate the IRS low sensitivity to silicate minerals like quartz and felspars. As an X-ray fluorescence spectrometer, PIXL will study at a scale of 100~$\mu$m chemical variation among the mineralogy detected at the millimeter scale by the IRS. At this scale, atomic substitution within minerals, grains and minerals assemblages, will provide important clues to understand weathering processes, especially bio-mediated processes, and potential for biosignature preservation in rocks \citep{Allwood2020}.
            
            To investigate atmospheric conditions and processes and the Martian climate, the IRS will work closely with the meteorological package MEDA as well as with imaging systems like Mastcam-Z and ECAM \citep{Maki2020}. MEDA will provide local measurement of the atmospheric humidity, that IRS investigation will complete by water column-density measurements. This will indicate how the boundary layer locally mixes with the free atmosphere. Mastcam-Z will also provide extremely accurate dust optical depth measurement with its direct solar imaging capability, which, associated with IRS observations of the forward-scattering peak (within 10$^{\circ}$ to 20$^{\circ}$ of the Sun) in the 1.3--2.6~$\mu$m, will provide tight constraints on the dust particle size and shape.
            
    \section{Instrument description}
    \label{Instrumentdescription}

        \subsection{Overview}
    
            The SuperCam instrument represents an advancement from the design of the ChemCam instrument operating onboard the Curiosity/MSL rover \citep{Wiens2012,Maurice2012}. This new instrument suite gathers four different remote-sensing techniques. In addition to the LIBS (Laser Induced Breakdown Spectroscopy) elemental analysis technique already implemented in ChemCam, a new Raman and time-resolved fluorescence spectroscopic analysis is implemented, as well as an Infrared passive Spectrometer. For context imaging, an improvement of the Remote Micro Imager (RMI) is provided by a new color detector. A microphone (MIC) has been added to record LIBS impacts, wind and rover sounds on the Martian surface \citep{Wiens2020,Maurice2020}.
    
            The SuperCam instrument consists of three separate major units: the Body Unit, the Mast Unit and the Calibration Targets. The three units are mechanically independent, simplifying interface controls as well as development overseas, under the leadership of the Los Alamos National Laboratory (LANL, USA). The following sections describe these units only to present information relevant to understand the SuperCam infrared spectrometer design, accommodation and operation. The reader is referred to \citet{Wiens2020,Maurice2020} for a complete presentation of the SuperCam instrument.
    
            \subsubsection{The Mast Unit}
        
                The Mast Unit (SCMU), provided by Institut de recherche en astrophysique et planétologie (IRAP, France) and funded by Centre national d'études spatiales (CNES) is extensively described in \citet{Maurice2020}. It consists of a Cassegrain telescope with a focusing stage, a “red” or “green” pulsed laser and its associated electronics, an infrared spectrometer, a color CMOS micro-imager, a microphone, and the associated electronics. The telescope ensures the emission and focusing functions for the laser but also the collection functions for the infrared spectrometer, the context imaging, and the UV and visible spectrometers located in the Body Unit. The telescope has the ability to focus from 1.1~m to infinity. When focused at 1.56~m, the laser has a focusing point on the secondary mirror. Firing the LIBS laser at this specific position is then forbidden; this focus position is called the "laser exclusion zone".
        
                The SCMU is composed of two parts: the optical box (OBOX) that hosts all the optical bench including the laser, the infrared spectrometer, the RMI and the microphone, and the electronic box (EBOX) that hosts all the electronic board (the Low Voltage Power Supply (LVPS) board for powering the instrument, the IRBOARD for driving the infrared spectrometer, the Laser Board for driving the laser, the Digital Processor Unit (DPU) board for managing and sequencing the overall instrument functions). The SCMU (OBOX + EBOX) dimensions are $167 \times 400 \times 219$~mm$^3$, for a total mass of 6.11~kg.
        
                As shown in Fig.~53 of \citet{Maurice2020}, the infrared spectrometer is located underneath the laser and electronic box within the Mast-Unit assembly.
        
            \subsubsection{The Body Unit}
        
                The SuperCam Body Unit (SCBU), provided by LANL and funded by NASA, is extensively described in \citet{Wiens2020}. It consists of three spectrometers covering the UV (245--340~nm), the violet (385--465~nm), and the visible and near-infrared (VNIR, 536--853~nm) ranges needed for the LIBS technique. The UV and violet spectrometers are Czerny-Turner units identical to those on ChemCam. The VNIR spectrometer uses a transmission grating and an intensifier so that it can be used for remote pulsed-laser Raman spectroscopy as well as LIBS and passive reflectance spectroscopy. The intensifier allows the short gating needed to remove the background light so that the weak Raman emission signals can be observed, as well as enabling time-resolved florescence studies. The Body Unit is responsible for the electrical and data interface with the Perseverance rover.
        
                A fiber optic cable, as well as signal and power cables, provided by the Jet Propulsion Laboratory (JPL), connect the SCBU and the SCMU.
        
            \subsubsection{The Calibration Target Unit}
        
                A set of calibration targets (SCCT), provided by the University of Valadolid  (Spain), will enable periodic calibration of the SuperCam instrument. These calibration targets are mounted on the deck of the rover \citep{Manrique2020,Cousin2020}, and especially comprise two targets devoted to in flight characterization and calibration of the SuperCam infrared spectrometer.

        \subsection{Requirements}
        
            The high level requirements for the infrared spectrometer are tabulated in Tab.~\ref{TabRequirementL5}. These requirements are derived from higher-level requirements set at the SuperCam level \citep{Wiens2020,Maurice2020} or at the Mars2020 level \citep{Farley2020}. These requirements were set and tuned to meet the SuperCam infrared spectrometer scientific objectives presented in Section~\ref{Scientificobjectives}, based on our previous knowledge of the Jezero mineralogy and of the Martian atmosphere near-infrared spectra.
   
            \begin{table}[]
                \centering
                \begin{tabular}{ |p{5.5cm}|c|p{4.8cm}| }
                    \hline
                    Parameter & Value & Comments \\
                    \hline
                    \hline
                    Wavelength Range & [1.3 -- 2.6]~$\mu$m  &\\
                    \hline
                    Spectral Resolution & 32~cm$^{-1}$ & 11.4~nm @ 1.95~$\mu$m \\
                    \hline
                    Resolution power & 170  & @1.90~$\mu$m\\
                    FOV & 1.15~mrad & \\
                    \hline
                    Spectral Sampling & 15~cm$^{-1}$ & \\
                    SNR & $> 56$ &\\ 
                    \hline
                    Number of spectral samples & 256 max & Adjustable from 1 to 256 \\
                    \hline
                    Acquisition mode & wvl scan & Adjustable among 256 \\
                    \hline
                \end{tabular}
                \caption{Mars2020 Level 5 requirements applying to the SuperCam infrared spectrometer.}
                \label{TabRequirementL5}
            \end{table}

            \subsubsection{Wavelength range}
            
                As detailed in Section~\ref{SecMineralIdentification} and Table~\ref{TabDetectionLimit}, most of the minerals whose identification constitutes the core of the SuperCam Infrared Spectrometer scientific objectives are detectable at wavelengths between 0.5 and 2.6~µm. The heritage of OMEGA and CRISM demonstrates that longer-wavelength (3.3--3.9~µm) corroboration is possible but not essential.  Some alteration products, biologically important like phosphates, nitrates, or perchlorates, also have near-infrared absorption features.
                
                However a trade-off had to made between the long-wavelength versus the short-wavelength cutoff, as covering the full 0.5--2.6~µm range was not compatible with the volume constraint of the Mast Unit. We decided to limit or wavelength range to 1.3--2.6~µm range for two reasons: i) the 0.5--1.3~µm range is covered by the combination of MastCam-Z and SuperCam Body Unit spectrometers, ii) the detection of phyllosilicate, carbonate and serpentines signatures up to 2.55~µm was a higher priority for a mission oriented on identififcation of ancient habitable environments and biosignatures than the search for iron-oxydes absorptions around 1.0~µm, most important for the identification of igneous minerals.
                
            \subsubsection{Spectral resolution and sampling}
            
\begin{figure*}
	                \centering
                    \includegraphics[width=15cm,clip=true,trim=15mm 5mm 15mm 140mm]{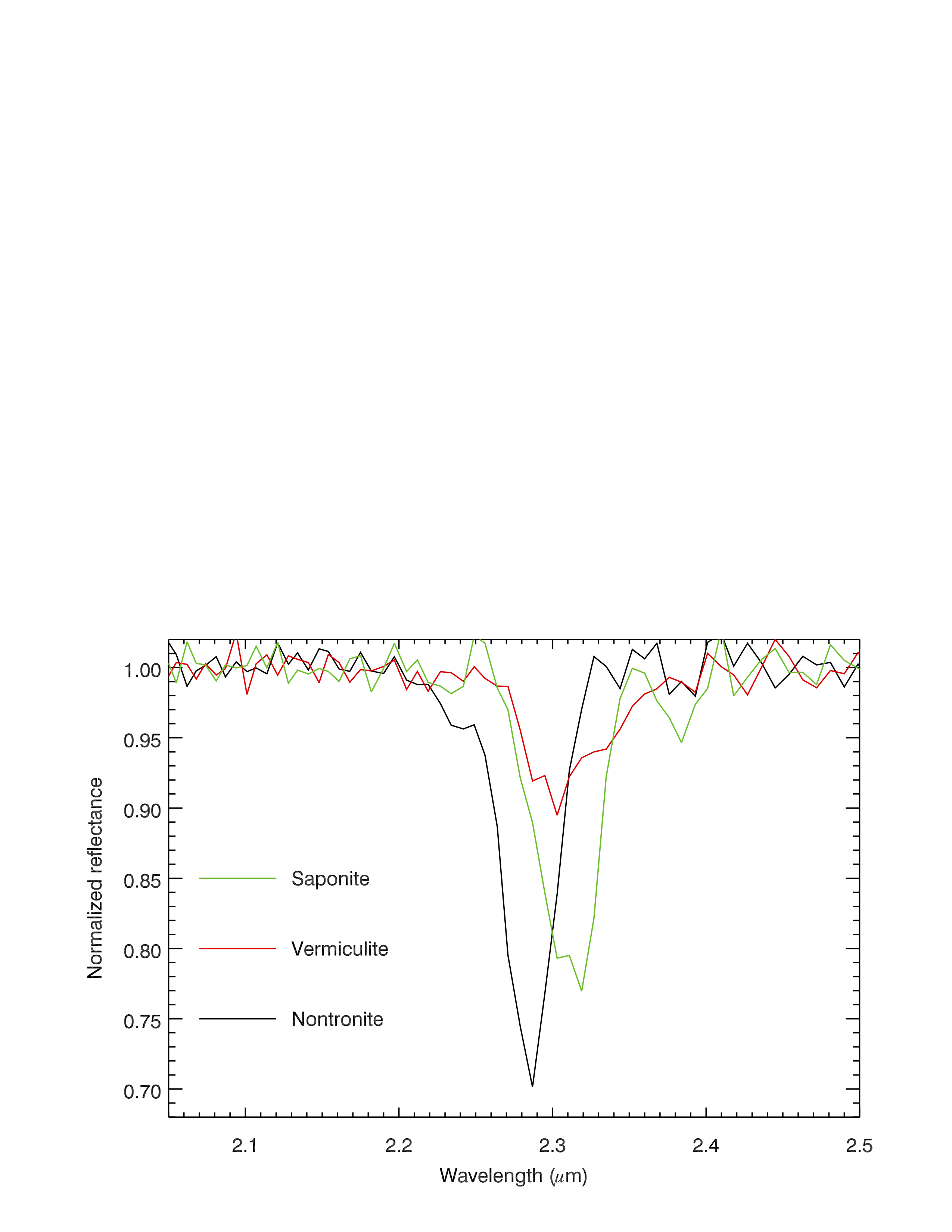}
                \caption{Examples of three reflectance spectra of phyllosilicates convolved and sampled to SuperCam IRS spectral resolution and sampling with a white noise added to in order to simulate the expected signal-to-noise ratio. Green line: saponite, a Mg-rich smectite (band at 2.32~$\mu$m), red line: vermiculite, a Fe/Mg smectite/mica (band at 2.30~$\mu$m), black line: nontronite, a Fe$^{3+}$ smectite (band at 2.28~$\mu$m).}
                \label{FigRequirements}
            \end{figure*}

                The ability of a spectrometer to identify mineral absorptions depends on its spectral resolution and its spectral sampling. We decided for a spectral sampling close to the Nyquist sampling, i.e. a spectral sampling close to half of the line spread function full (LSF) width half maximum (FWHM). \citet{Swayze2003} conducted a comprehensive study of the required spectral resolution needed to identify a wide range of minerals. Their study concluded that a spectral resolution of $11\pm2$ discriminates key minerals for a sampling step equal to the spectral resolution. We therefore set a requirement for a spectral resolution of 32~cm$^{-1}$, i.e. $11.4$~nm at the center of our spectral domain. Fig.~\ref{FigRequirements} shows that it allows us to discriminate absorptions of Fe-bearing and Mg-bearing phyllosilicates, key species to identify.
                
            \subsubsection{Signal to noise ratio}
            
                \citet{Swayze2003} showed that a signal-to-noise ratio (SNR) of about a 100 allows to identify key-minerals. The study of \citet{Swayze2003} was conducted for space-borne spectro-imaging spectrometers that usually operate at optimal solar insolation close to local noon. An infrared spectrometer on a rover will operate in more diverse illumination conditions, from local morning to local afternoon, and target the rocks available in the workspace that maybe in partially or totally in shadow. It will also encounter a wider range operating temperatures than an orbiter. We therefore set a SNR of 60, but for intermediate solar illumination conditions, 300~W/m$^2$, and rocks of an albedo of 0.3, for operating temperatures of up to $-5$\textdegree{C}. Such a requirement will result in a SNR in excess of 100 for optimal solar insolation of 600~W/m$^2$ comparable to that achieved by space-borne platforms.
                
                Given the Mast Unit optical transmission in the infrared, which is imposed by the Raman and LIBS spectroscopy functions, this required SNR was achievable only with an FOV of 1.15~mrad. This is larger, but still comparable with the SuperCam FOV in the visible domain, 0.76~mrad.
                
                This requirement on the signal to noise ratio also impacted the requirements on the radiometric calibration. In order to take full advantage of our SNR in detecting weak mineralogical features, the relative calibration on adjacent wavelengths was set to be better than 1\%.

        \subsection{Design overview}
   
\begin{figure*}
	            \centering
                \includegraphics[width=15cm]{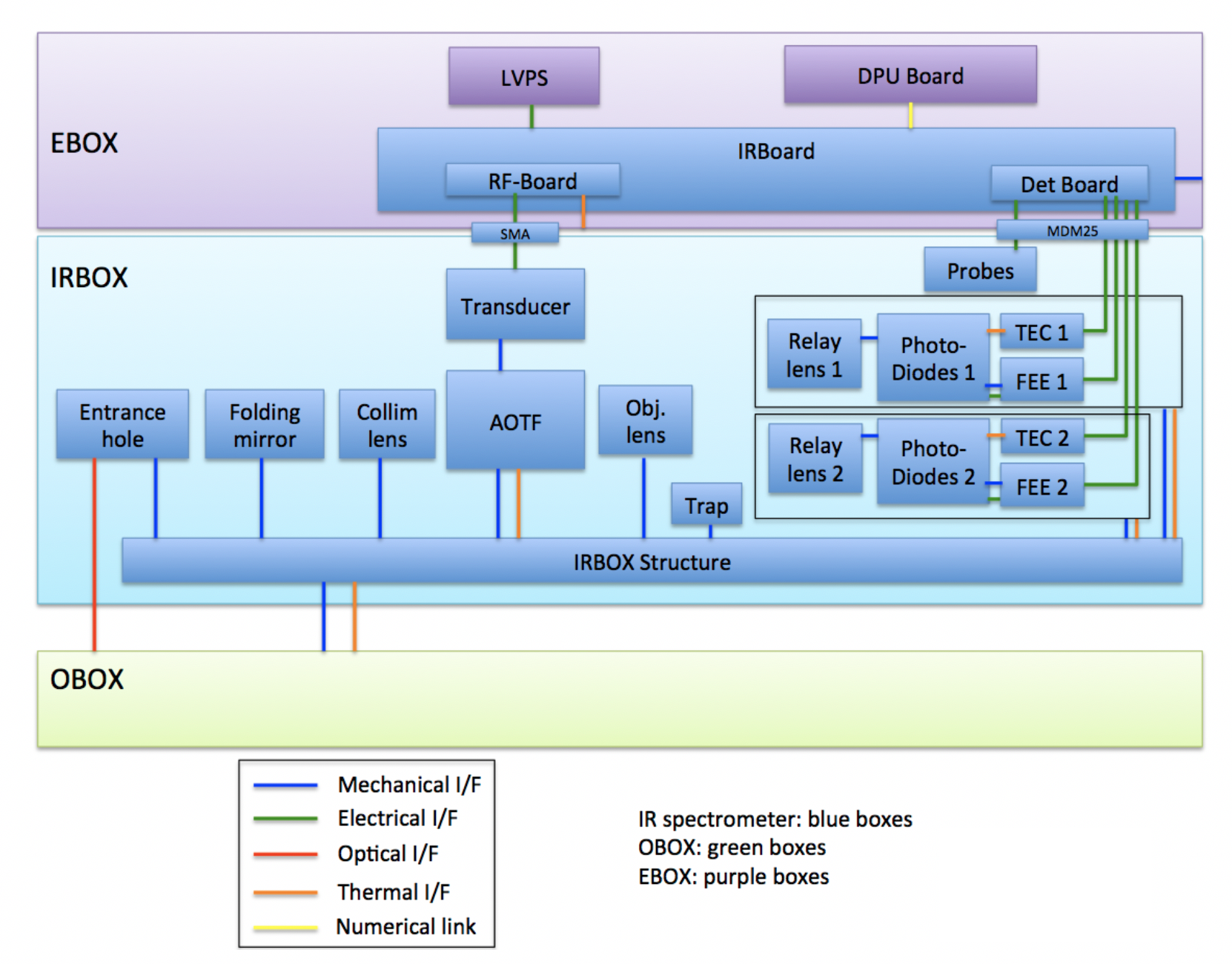}
                \caption{The interface block diagram of the SuperCam infrared spectrometer.}
                \label{FigBlockdiagram}
            \end{figure*}

            The entire SuperCam infrared spectrometer and its interfaces with the SCMU are described in the block diagram displayed in Fig.~\ref{FigBlockdiagram}. The SuperCam infrared spectrometer is composed of:
            \begin{itemize}
                \item The IRBOX that contains the spectrometer itself, physically located in the OBOX. The IRBOX includes all the opto-mechanical and detection functions of the spectrometer. It is mechanically and optically linked to the OBOX;
                \item The IRBoard, which is physically located in the EBOX. The IRBoard drives the AOTF and the detectors. It is directly linked to the SCMU-DPU board. 
            \end{itemize}

            The SuperCam infrared spectrometer is located in the collection path of the telescope. At the output of the Cassegrain telescope a set of dichroic mirrors split the collected bandpass. In the IR path, three lenses are used to:
            \begin{itemize}
                \item Inject the image in a 400~$\mu$m diameter entrance hole with a 0.18NA;
                \item Relay the entrance pupil in the spectrometer near the entrance of the AOTF (see Sec.~\ref{SecOpticalConcept}).
            \end{itemize}
    
            \subsubsection{General optical concept}
            \label{SecOpticalConcept}
\begin{figure*}
                    \centering
                    \includegraphics[width=15cm]{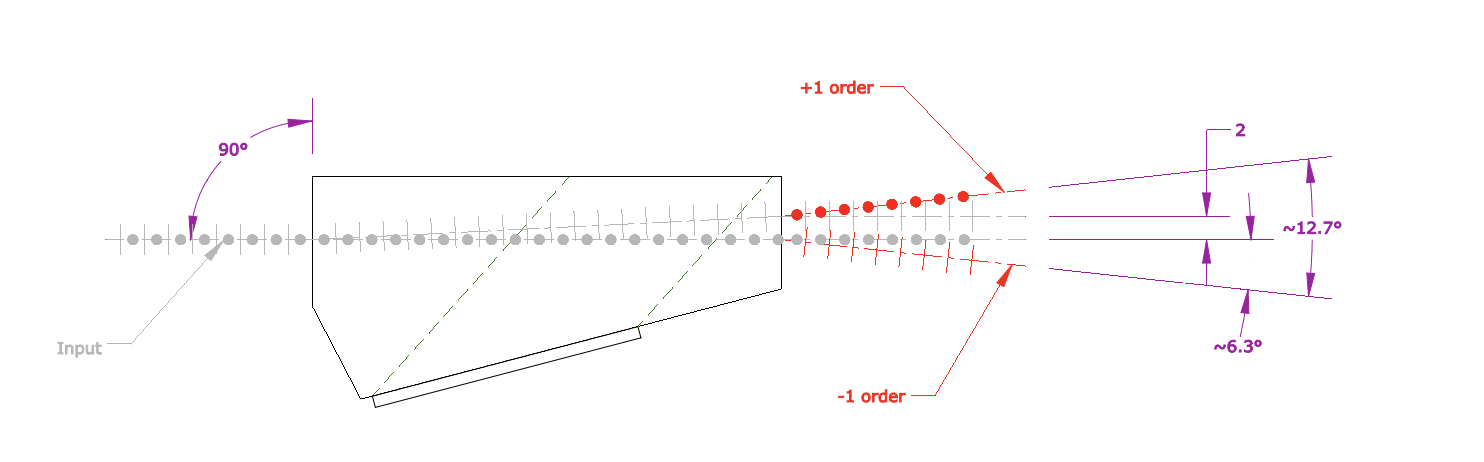}
                    \caption{Schematic diagram presenting the AOTF diffraction principle. The $+1$-order stands for the o-order, while the $-1$-order stands for the e-order. The beam labelled 2 represents the undiffracted beam. \copyright Gooch \& Housego.}
                    \label{AOTF_principle}
                \end{figure*}

                The SuperCam infrared spectrometer concept is based on the scan of the full spectral range by an Acousto-Optic Tunable Filter (AOTF). It capitalizes on the successful use of an AOTF within the SPICAM instrument \citep{Bertaux2000,Korablev2006}. An AOTF relies on the diffraction of incident light by non-standing acoustic waves in a tellurium dioxide (TeO$_2$2) birefringent crystal generated by an electro-acoustic transducer. In the frame of the Bragg’s diffraction formalism, the acoustic waves stream behaves as a thick grating for which the phase matching condition leads to the emergence of an unique diffracted monochromatic light beam deviated from the non-diffracted polychromatic beam. The central wavelength of this beam is bijectively determined by the acoustic waves frequency (itself determined by the radio frequency (RF) signal supplying the transducer), and its spectral width is related to the crystal’s geometry. Moreover, due to TeO$_2$ birefringence, a second couple of diffracted and non diffracted beams is symmetrically generated to the optical axis. Both couples of beams, corresponding to one extraordinary ray (e-ray order) and one ordinary ray (o-ray order), have the same properties excepted they are linearly cross-polarized. Figure~\ref{AOTF_principle} exhibits the concept of an AOTF with the two polarized output at a given wavelength.

                In the SuperCam infrared spectrometer, the zero-order is trapped in a light trap. The e-ray and the o-ray orders are projected on two different photodiodes by using two lenses. The optical path inside the SuperCam infrared spectrometer consists in:
                \begin{itemize}
                    \item An entrance hole illuminated by the MU-telescope that determines the FOV of the spectrometer;
                    \item A folding mirror and a single-ZnSe collimator lens;
                    \item The AOTF;
                    \item A single-ZnSe objective lens that images the 3-AOTF outputs. While the zero-order is trapped in the objective image plane, the e-ray and o-ray orders pass through;
                    \item Two photodiodes, each of them mounted with a single ZnSe relay lens that images the entrance hole on the sensitive part of the photodiode. The e-ray path is folded in front of the photodiode.
                \end{itemize}
        
\begin{figure*}
	                \centering
	                \includegraphics[scale=.50]{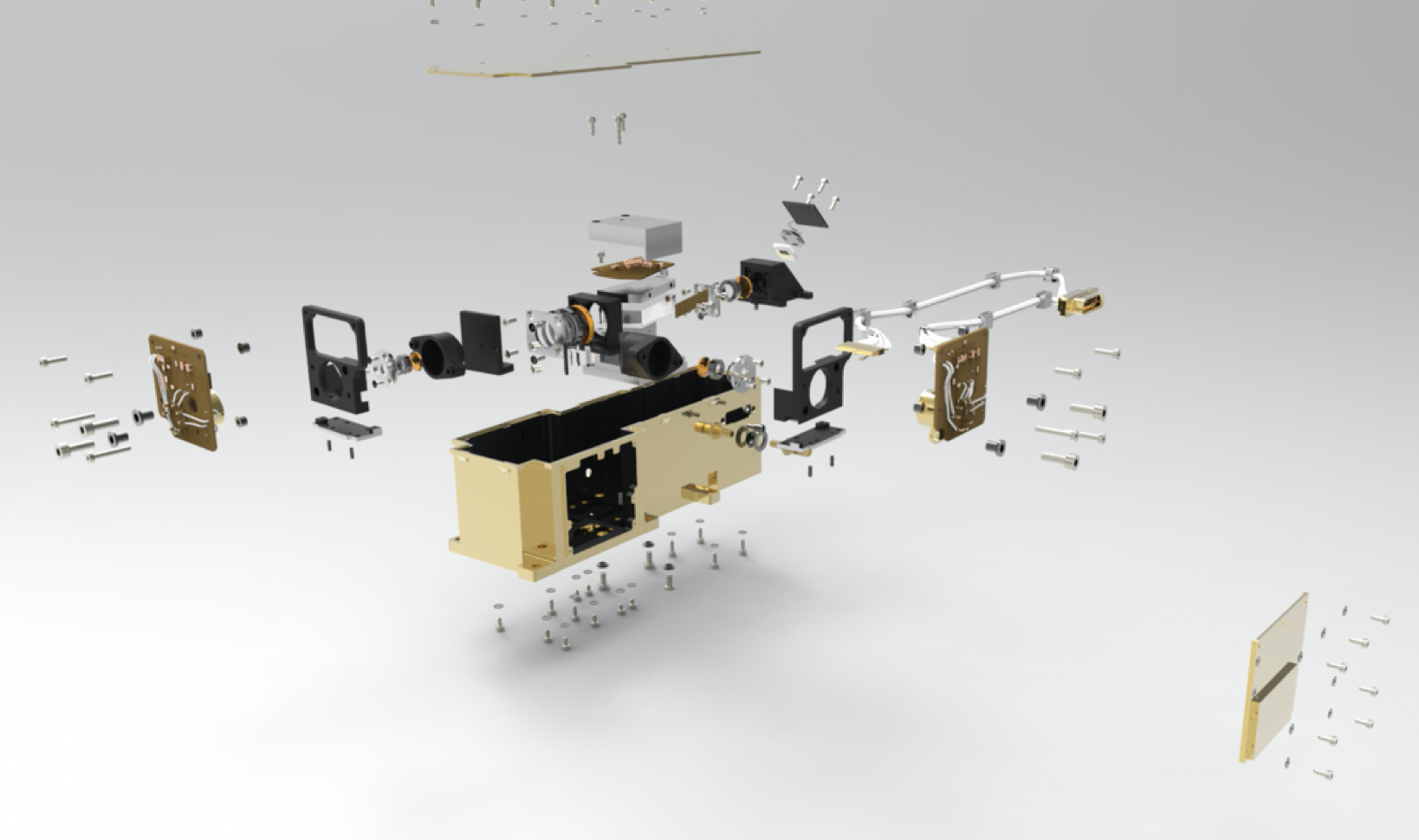}
	                \includegraphics[scale=.70]{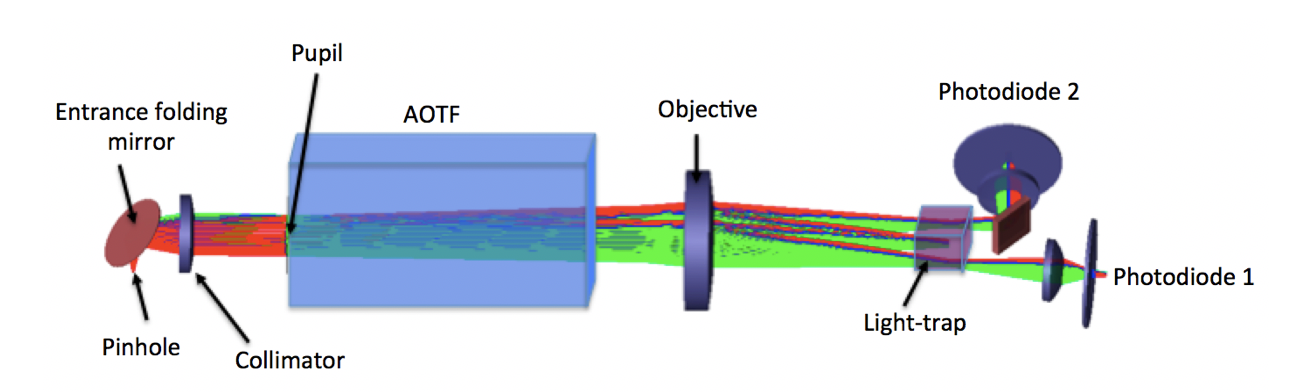}
	                \caption{Upper planel: Exploded view of the SuperCam infrared spectrometer, including all its mechanical, optical, and electrical parts and pieces. Lower panel: optical design of the SuperCam infrared spectrometer.}
	                \label{IRS_view}
                \end{figure*}

                Figure~\ref{IRS_view} shows the different mechanical, optical and electrical elements of the spectrometer.
            
            \subsubsection{The AOTF}
       
                The AOTF constitutes the heart of the instrument. It has been designed and manufactured by the Gooch \& Housego company, Illminster, UK. It weights 84g.
       
                The SuperCam AOTF was designed to stabilize the two polarized outputs at $\pm3$~mrad for the whole RF frequency range. The angular separation between the diffracted orders and the zero-order was set to be higher than 6$^{\circ}$. The zero-order, comprising most of the input flux, in order to be optically trapped before the detection to reduce straylight within the IRBOX, had to deviate from less than 0.5$^{\circ}$ from the input beam. The maximum total acceptance angle for the input beam was set at 4$^{\circ}$.
       
                The AOTF has been intensively tested in representative environments to verify that all these requirements were met.
            
            \subsubsection{Detection}
            \label{SecDetection}
            
                The detection is performed by two MCT(HgCdTe) photodiodes provided by the Judson company (USA, formerly Teledyne), and mounted with a front-end electronics (FEE). The J19 photodiode series equivalent circuit is a photon-generated source with parallel capacitance, shunt resistance and series resistances. The part number is J19TE3:2.8-66C-R01M with a T066 package. The active size diameter is 1mm with a 50\% cut-off at 2.8~$\mu$m and a peak minimum responsivity of 1.3A/W at 2.6~$\mu$m. An anti-reflection coated sapphire window ensures the sealing of the package.
            
                In order to limit the dark current, the photodiodes are cooled down using a 3-stage Thermal Electric Cooler (TEC, Peltier effect thermal device) provided with the photodiode package. Only one photodiode and its associated FEE can be operated at a given time. The e-ray photodiode and its FEE were accommodated as a cold redundancy only. They will be operated only in case of failure of the o-ray photodiode.
            
                \paragraph{Detector signal \& detector current level.}
            
\begin{figure}
	                    \centering
		                \includegraphics[scale=.27]{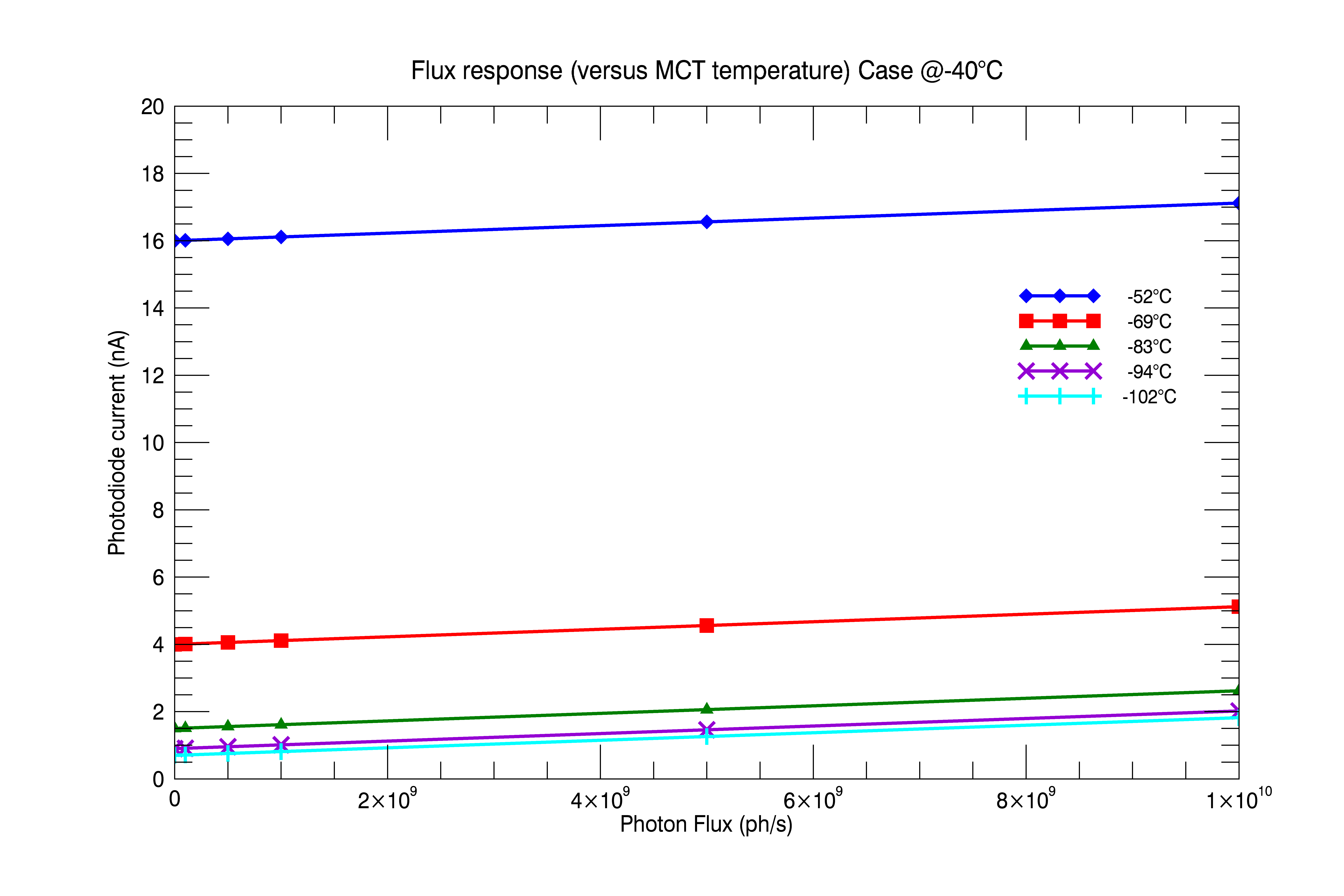}
	                    \caption{Photodiode currents versus photon flux for various MCT temperatures. These estimates were obtained by measuring the dark and readout currents in a thermal vacuum chamber set at $-40$°C, AOTF off, and taking into account the nominal photodiode sensitivity.)}
	                    \label{FluxResponse}
                    \end{figure}

                    The detector signal is read in the form of an electrical current generated by the photodiode as a function of i) the target IR flux, ii) the spectrometer thermal background, and iii) the photodiode dark current. The photodiode dark current is the dominant current, ranging between 40~nA, for a detector cold face temperature of $-60^{\circ}$C, and $\sim2$~nA at $-90^{\circ}$C (Fig.~\ref{FluxResponse}). The thermal background current is the second contribution to the total current, with an amplitude of a few nA at a spectrometer temperature of $-30$\textdegree{C}. Finally, the target IR flux current ranges between 10~pA and 1~nA, depending on the insolation and albedo of the target.

                    Since, the current generated by the target IR flux is not the dominant contribution to the total current, the detector readout stands as the most critical acquisition of all the IRS measurements. This situation has a strong impact on the design of the instrument, but also on its acquisition strategy, which we explain in more details in the following paragraph \textit{Detector signal acquisition strategy}.
                    
                    Regarding the instrument design, first the spectrometer temperature was stabilized by adequate thermal contacts to dump the spectrometer heat into the OBOX, therefore increasing the thermal inertia of the IRBOX. Hence, during an acquisition, the thermal background variation remains lower than a tenth of a pA. Then, we also designed the detection chain with the aim of reducing the readout noise on the photodiode electrical current. To do so, the FEE has been designed to be physically located as close as possible to the photodiode, in order to integrate the photodiode current and to pre-amplify the signal while minimizing electromagnetic interference. Finally, the photodiode reverse bias voltage was set at 100~mV, the trade-off voltage recommended by the manufacturer.

                \paragraph{Detector signal acquisition strategy}

                    The acquisition strategy is designed to subtract the environmentally generated currents from the total current in order to determine the observation IR current. To achieve this objective, the acquisition strategy interlaces a measurement of the scientific signal (AOTF ON) and of the background signal (AOTF OFF) for each RF frequency. The difference between these two acquisitions, the signal and dark currents, represents the diffracted science IR flux as well as possible thermal background variations between the two acquisitions. In order to avoid large thermal variations between two consecutive measurements that could bias the science measurement, the design paid great attention to the thermal stability of the instrument (see section~\ref{Section_Thermo-mechanical}).
                    
            \subsubsection{Front End Electronics}
            \label{SectionFrontEndElectronics}

\begin{figure}
	               \centering
	               \includegraphics[scale=.35]{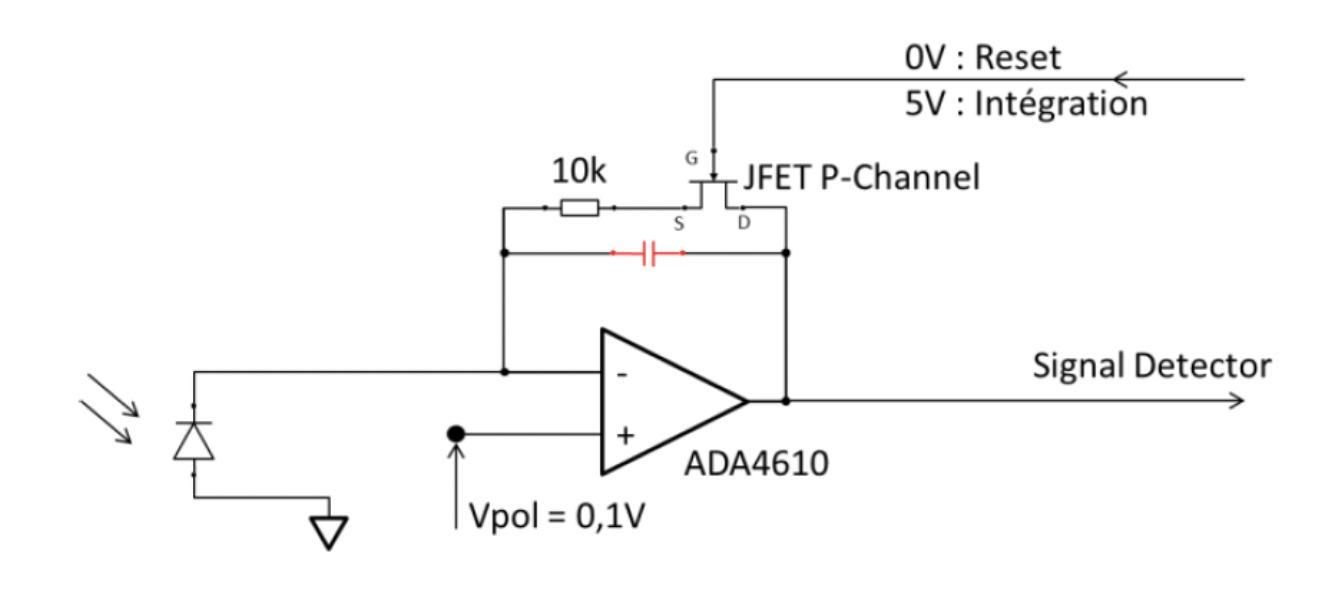}
	               \caption{The FEE integrator electronic circuit.}
	               \label{Integrator}
                \end{figure}

                The current signal processing chain is mainly composed of a transimpedance pre-amplifier. The readout current pre-amplifier is an integrator amplifier that sums all the currents generated by the photodiode into a capacitance during the defined integration time. After acquisition, the integrator is reset before the following integration. The J-FET inputs OP-Amp ADA4610-S by Analog Devices was selected for its low noise and low input current features. The reset switch is performed with a discrete SMD J-FET transistor P-Channel instead of an IC component to save space on the small FEE package, while reaching similar performance to those of an IC component (Fig.~\ref{Integrator}). 

                In this design, the voltage across the capacitance represents the charged average current during a single integration time. The voltage depends on the capacitance value and on the integration time through the relation:
                    
                \begin{equation*}
                        V_{int}=\frac{I_{diode}+T_{int}}{C_{int}}+V_{0}
                \end{equation*}
                where $V_{int}$ is the integrator output voltage, $I_{diode}$ the average current generated by the photodiode during the whole integration time, $T_{int}$ the integration time, $C_{int}$ the integration capacitance value, $V_{0}$ the offset voltage at the beginning of the integration.
                    
                In this design, increasing the integration time tends to improve the signal-to-noise ratio (SNR). However, the integration time should be kept short enough to avoid important current variations during two consecutive acquisitions (between signal acquisition and dark acquisition) due to any thermal and environmental background variations. The integration time should also be limited to avoid the saturation of the amplifier output (maximum output voltage of $+5$V). The integration time is optimized to adapt the integrator gain to the current level as a function of the photodiode temperature. The SNR is also increased by the digital sample accumulation implemented on the DPU driver of the SuperCam infrared spectrometer by repeating the measurement several times and by adding up the results.
                    
                The offset voltage $V_0$ is due to the current discharge generated at the junction capacitance in the analog switch at its opening (i.e., at the start of the integration). This current discharge charges down the integration capacitance under reset voltage by some tenths of a volt according to the integration capacitance value. This offset voltage is difficult to calculate and simulate, and is suppressed by subtraction of the two consecutive signal and dark acquisitions obtained with same offset voltage.
                    
            \subsubsection{Thermo-mechanical concept and design}
            \label{Section_Thermo-mechanical}
        
                The IRBOX baseplate, manufactured in house at the Laboratoire d'études spatiales et d'instrumentation en astrophysique (LESIA, Observatoire de Paris), is gold coated externally to minimize radiative thermal coupling with the environment. The baseplate is internally blackened (PNC coating) to limit internal straylight. All the optics are mounted by means of barrels to the baseplate. The AOTF is mounted on the baseplate with a Choterm layer in order to evacuate the heat generated by the acoustic wave into the baseplate.
        
                Each photodiode is mounted on a bracket that also accommodates the associated FEE. The heat generated by the TEC is also conducted into the baseplate for dissipation. The baseplate is mounted on the OBOX in order to directly dissipate the heat generated within the IRBOX into the full OBOX structure. The power dissipated into the OBOX structure can reach 3~W for the AOTF at peak consumption, and a maximum of 1.5~W max for the operating photodiode, adding to a maximum total of 4.5~W. This thermal coupling between the IRBOX mechanical structure and the OBOX is set to maintain the spectrometer at the coolest possible temperature, and stabilize its temperature to maximize the SNR.
        
            \subsubsection{Infrared electronic board}
        
                The IRBoard, located in the EBOX, drives the SuperCam infrared spectrometer and implements three main functions:
                \begin{itemize}
                    \item it generates the RF signal to drive the AOTF filter;
                    \item it drives the signal acquisition function for the two detectors in parallel with the housekeeping signals;
                    \item it regulates the TEC for detectors cooling.
                \end{itemize}

                The acquisition function is further physically separated into two different parts: 
                \begin{itemize}
                    \item the acquisition part implemented on the IRBoard that contains the signal processing and the digital conversion functions;
                    \item the FEE, implemented at the rear of each detector inside the IRBox, which pre-amplify the detector signals (see section~\ref{SectionFrontEndElectronics}).
                \end{itemize}

\begin{figure*}
	                \centering
		            \includegraphics[scale=.45]{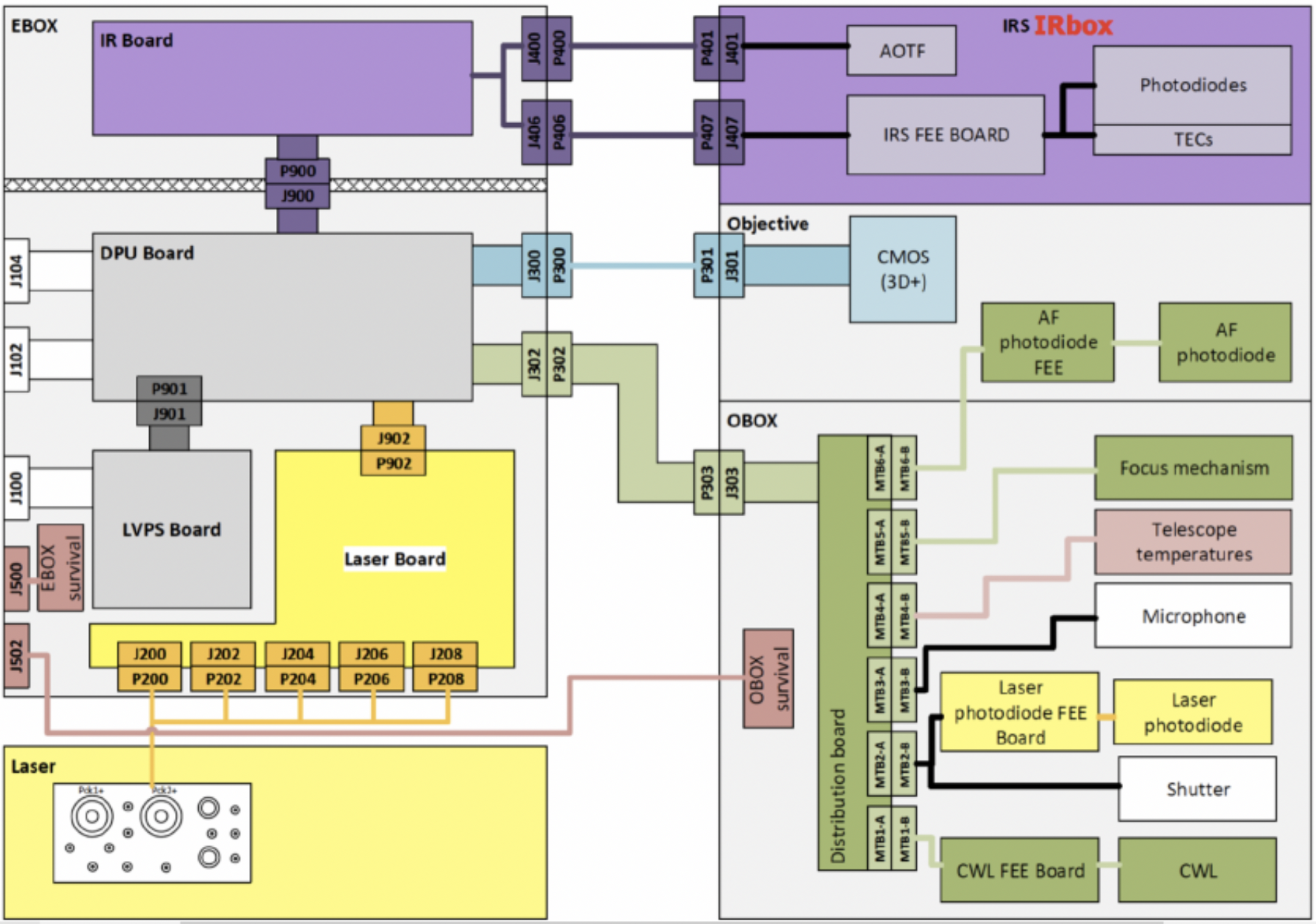}
	                \caption{Schematic of the SCMU general electrical architecture and interface to the IRBOX and IRBoard (purple).}
	                \label{ElecArchitecture}
                \end{figure*}

                The Supercam DPU Board provides to the IR Board the requested command signals and power supply voltages (generated first by the LVPS Board \citep{Maurice2020}). The different functions are power-supplied by the following dedicated voltages: 
                \begin{itemize}
                    \item $\pm5$V and +20V power the RF functions;
                    \item $\pm5$V are used for the signal acquisition function;
                    \item $+3.3$V is dedicated to the TEC regulation function.
                \end{itemize}
                These supply voltages are filtered by common mode filters and differential filters. Optocouplers interface the command signals, generated by a FPGA located whithin the DPU Board, to the FEE and TEC photodiodes located within the IRBOX (Fig.~\ref{ElecArchitecture}).

                \paragraph{RF generation function}

                    \begin{figure*}
	                    \centering
		                \includegraphics[scale=.50]{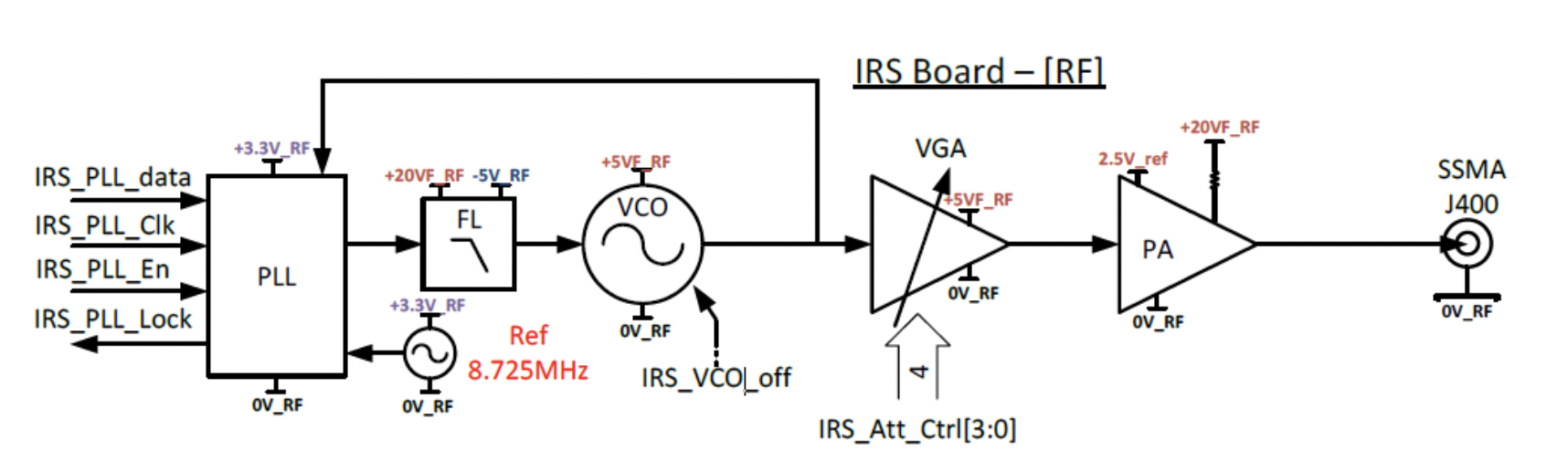}
	                    \caption{Schematic of the SuperCam infrared spectrometer RF generation circuit.}
	                    \label{RFGenerator}
                    \end{figure*}

                    The DPU board configures the Phase Locked Loop (PLL) with a specific Serial Peripheral Interface (SPI) bus in order to lock the Voltage Controlled Oscillator (VCO) at the requested frequency within a bandwidth from 33.8~MHz to 68.7~MHz, with 256 steps of 136.33kHz, corresponding to the AOTF IR 1.3--2.6 $\mu$m spectral bandpass. The scan is performed by increasing the RF frequencies, so the spectrum is acquired from long wavelengths to short wavelengths. While the loop is locked, the “Lock Detect” signal returns a “high” state to the DPU board. The RF signal generated by the VCO is split into two components: a first component is transmitted back to the PLL for close-loop control, while the second component feeds in the amplification and transmission stages.
    
                    The Variable Gain Amplifier (VGA) is the first amplification stage. It allows the IRBoard to set the power level of the RF signal by changing the gain with a four-bit command signal configured by the DPU board. The Power Amplifier (PA) is the final stage of the RF channel. It amplifies the RF signal up to $+35$~dBm ($3$~W). The VCO can be disabled in order to switch off the RF signal generation. The RF output signal is transmitted to the AOTF (located inside the IRBox) with a SSMA connector interface (Fig.~\ref{RFGenerator}).

                \paragraph{Acquisition function}

                    The SuperCam infrared spectrometer needs to acquire several types of signals:
                    \begin{itemize}
                        \item the detector signal: the readout current of the two photodiodes (2 channels);
                        \item the detector temperature signal: the temperature of photodiode used for regulation and science correction,
                        \item three environmental temperature signals: the temperatures of the IRBoard, AOTF and detector hot case,
                        \item the RF Power signal, in the form of the drain current of the Power Amplifier.
                    \end{itemize}

                    \begin{figure*}
	                    \centering
		                \includegraphics[scale=.70]{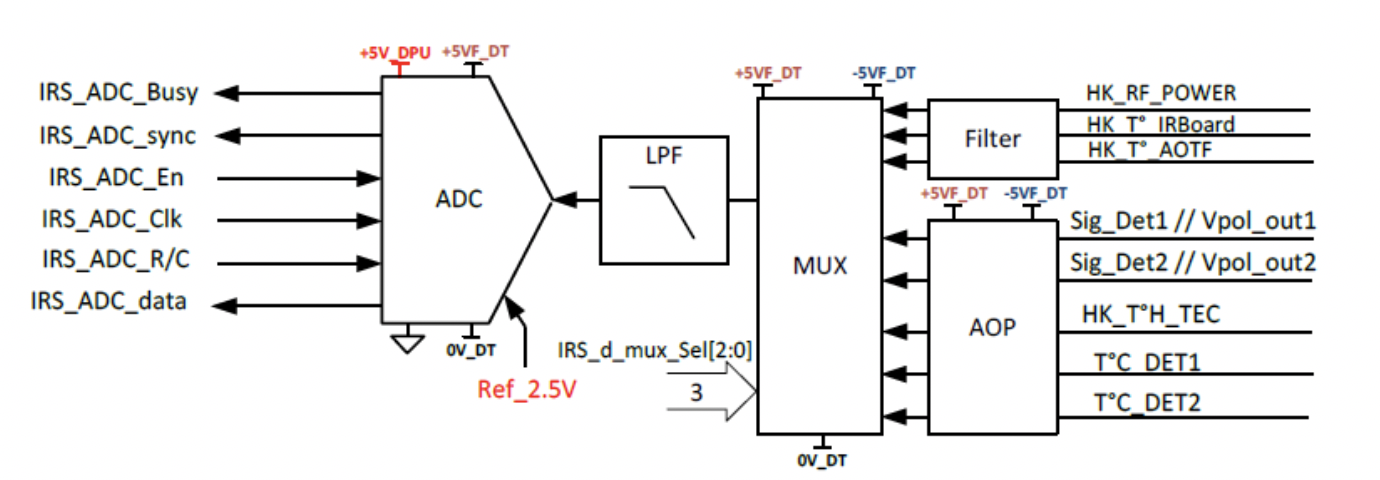}
	                    \caption{Schematic of the SuperCam infrared spectrometer Acquisition Circuit.}
	                    \label{AcquisitionCircuit}
                    \end{figure*}

                    Each signal is selected by 8 to 1 analog multiplexers to be sampled by the 16-bits Analog to Digital Converter (ADC) through low pass filters. Both the ADC component and the multiplexer component are digitally controlled by the DPU board. Operational Amplifiers are used to translate the different signal voltage ranges to match the ADC input voltage range lying between 0~V to 5~V. Nominally, only one detector can be activated (cooled and polarized) during operations. As a result, the IRBoard is designed to acquire only the selected detector and temperature signals along with the “RF power signal” during a typical spectrum acquisition. The HK Temperature signals are acquired before and after a full spectrum acquisition (Fig.~\ref{AcquisitionCircuit}).

                \paragraph{Thermo-electric Cooler temperature regulator}

                    The Judson J19 photodiode package contains a TEC and a Thermistor used for the closed-loop control of the photodiode photosensitive surface temperature. A digital proportional–integral–derivative (PID) controller is implemented in the FPGA located whitin the DPU board. This controller uses the detector temperature measurement to control the supply current to the TEC in order to stabilize the detector temperature at the configured set point. The thermistor is a Negative Temperature Coefficient (NTC) sensor that displays typical non-linear characteristics.

                    To reduce the non-linearity and increase the precision of the NTC sensor around the defined temperature, a voltage bridge divider is used to convert the resistance value into a voltage signal. The second resistor used in the bridge divider with thermistor sets the value of the thermistor at the center of desired temperature range. The TEC current is regulated by a P-MOS low logic level transistor, used as a very low noise serial ballast current driver. This solution was chosen to reduce any noise disturbances possibly transmitted to the detector.

                    The P-MOS transistor is controlled by a Pulse Width Modulation (PWM) signal generated by the FPGA located in the DPU board, and is lowpass filtered to reject high frequency signals, and to keep only the DC part, as an image of the PWM duty cycle. The duty factor of the PWM allows the IRBoard to set and to control the  current flowing through the P-MOS transistor supplying also the TEC inside the photodetector package.

            \subsubsection{Instrumental probes}
            \label{SectionHK}

                Several probes allow control of the instrument temperatures, voltages, and signals and help in checking the instrument health status. The measurements are recorded in ADU and communicated to the rover and eventually sent back to Earth along with the scientific data in the datareply. Table~\ref{TabHK} summarizes the different housekeeping (HK) measurements and parameters sent back to Earth. The probe signals must be converted into physical units through dedicated and calibrated transfer functions, which are presented in the Appendix~\ref{AppendixHK}.

                \begin{table}[]
                    \centering
                    \begin{tabular}{c|c}
                        Returned HK Data & Types  \\
                        \hline
                        \hline
                        AOTF Temperature & Measurement \\
                        TEC hot face Temperature & Measurement\\
                        TEC cold face Temperature & Measurement \\
                        TEC cold face setpoint & Copy of the command \\
                        IRS 20V Power-voltage & Measurement \\
                        FEE 3V Power-Voltage & Measurement \\
                        AOTF $\pm$5V Power-Voltage & Measurement \\
                        TEC $\pm$ 8V Power-Voltage & Measurement \\
                        PLL Lock & Measurement \\
                        TEC Ready & TEC temperature within 1K of the setpoint \\
                        Spectel Saturation & Detector read-out larger 12000 ADU
                    \end{tabular}
                    \caption{List of housekeeping (HK) data that are returned in the data reply with each single spectrum}
                    \label{TabHK}
                \end{table}

        \subsection{Instrument budget}

            At the IRS level the power line consumption is listed in Tab.~\ref{Tablepowerconsumption}

            \begin{table}[]
                \centering
                \begin{tabular}{c|c}
                    Line & Current  \\
                    \hline
                    \hline
                    $-5$~V RF & 6.3~mA \\
                    $+5$~V RF & 142.5~mA \\
                    3.3~V & 4.1~mA \\
                    20~V & 21.3~mA \\
                    $+8$~V DT & 51.1~mA \\
                    $-8$~V DT & 17.7~mA
                \end{tabular}
                \caption{Power consumption lines at the IRS level.}
                \label{Tablepowerconsumption}
            \end{table}

            The IRBOX mass is 395~g.

        \subsection{Instrument operating modes}
    
            The SuperCam infrared spectrometer is operated in a unique set of acquisition modes addressed by the instrument command \textit{TakeIRSpectrum}.

            \subsubsection{Spectrum tables}
        
                Depending on the instrument temperature, the acquisition of a single spectel can take up to 100~ms. In this context, operation on Mars will trade off between the number of spectra acquired and the number of spectels sampled in each individual spectrum. In addition, the uplink volume between the Earth and the surface of Mars is constrained, and the Mars2020 project did not accommodate for the 256 bytes per spectrum needed to select the 256 spectels independently.
            
                For these reasons, it was decided that the IRS spectels can only be addressed by 16 different tables, designed in advance and stored in the SCMU DPU. These tables specify the spectels that will be observed, and $N_\text{acc}$, the number of successive pairs of AOTF ON and AOTF OFF measurements performed for each individual spectel.
            
                The 16 different tables stored in the SCMU DPU are presented in Tab.~\ref{NaccTable}. Besides table \#4 that commands all spectels, tables \#2 and \#3 were designed to obtain a full spectrum at half the nominal sampling rate and can be merged to create a full spectrum without hitting the instrument timeout (see Sec.~\ref{SecTimeOut}). Tables \#0 and \#1 are specifically designed to optimize the total exposure time, while sampling the most important mineralogical bands at $\sim1.4$~$\mu$m, $\sim1.9$~$\mu$m and between 2.1 and 2.6~$\mu$m. Compared to table\#0, table \#1 samples the 2.0-$\mu$m CO$_2$ atmospheric band in greater details to allow for a more precise removal of atmospheric features. Several small tables, comprising a handful of spectels, target individually the most important mineral bands diagnostic of hydration or CO-stretching. These small tables are designed to identify some specific minerals in a wide scene using the scan mode. Finally, two tables (\#14 \& \#15) are designed to survey gaseous absorptions (CO$_2$, H$_2$O, and CO), as well as dust and water ice opacities. These two tables will be used either in passive sky observations or by observing the illuminated white calibration target.
                
\begin{figure*}
                    \centering
                    \includegraphics[width=0.85\linewidth, clip=true, trim=0 70mm 0 65mm]{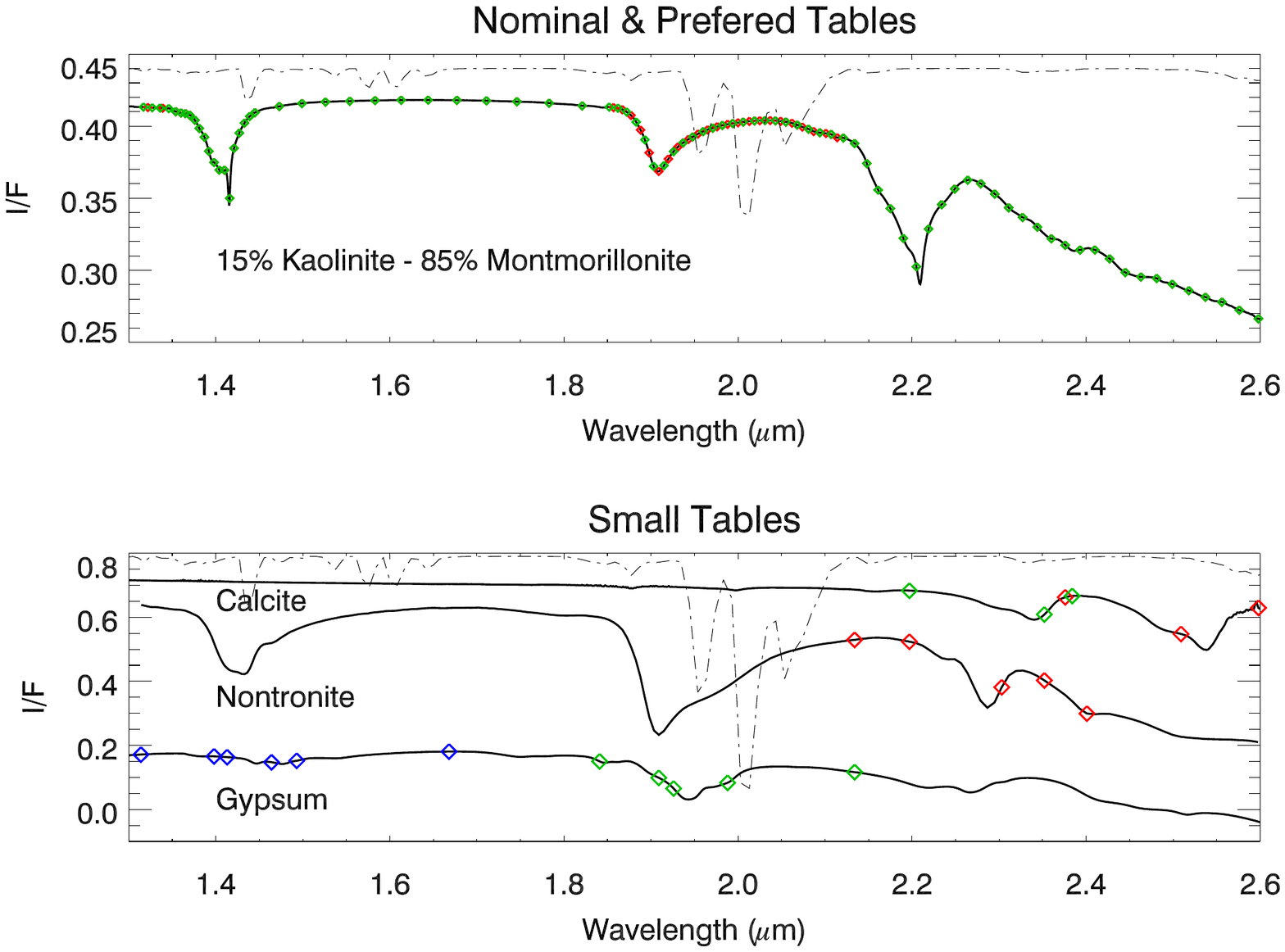}
                    \caption{Wavelenghths sampled by some of the currently loaded spectrum tables overplotted on mineralogical spectra. Upper panel: the nominal (green diamonds) and the preferred (green \& red diamonds) tables overplotted on a spectrum of Kaolinite and Montmorillonite. Lower panel: the BD2p50 (red diamonds) and the BD2p34 tables (green diamonds) overplotted on a calcite spectrum, the BD2p3 table (red diamonds) overplotted on a nontronite spectrum (red diamonds), and the BD1p9 Full (green diamonds) and the BD1p4-1p5 Full tables (blue diamonds) on a Gypsum spectrum. The gypsum spectrum is shifted by $-0.3$. A Mars atmosphere transmission spectrum, scaled to each figure, is overplotted in dash-dotted line.}
                    \label{FigNaccTables}
                \end{figure*}

\begin{figure}
                    \centering
                    \includegraphics[width=0.85\linewidth,clip=true,trim=0mm 80mm 0mm 80mm]{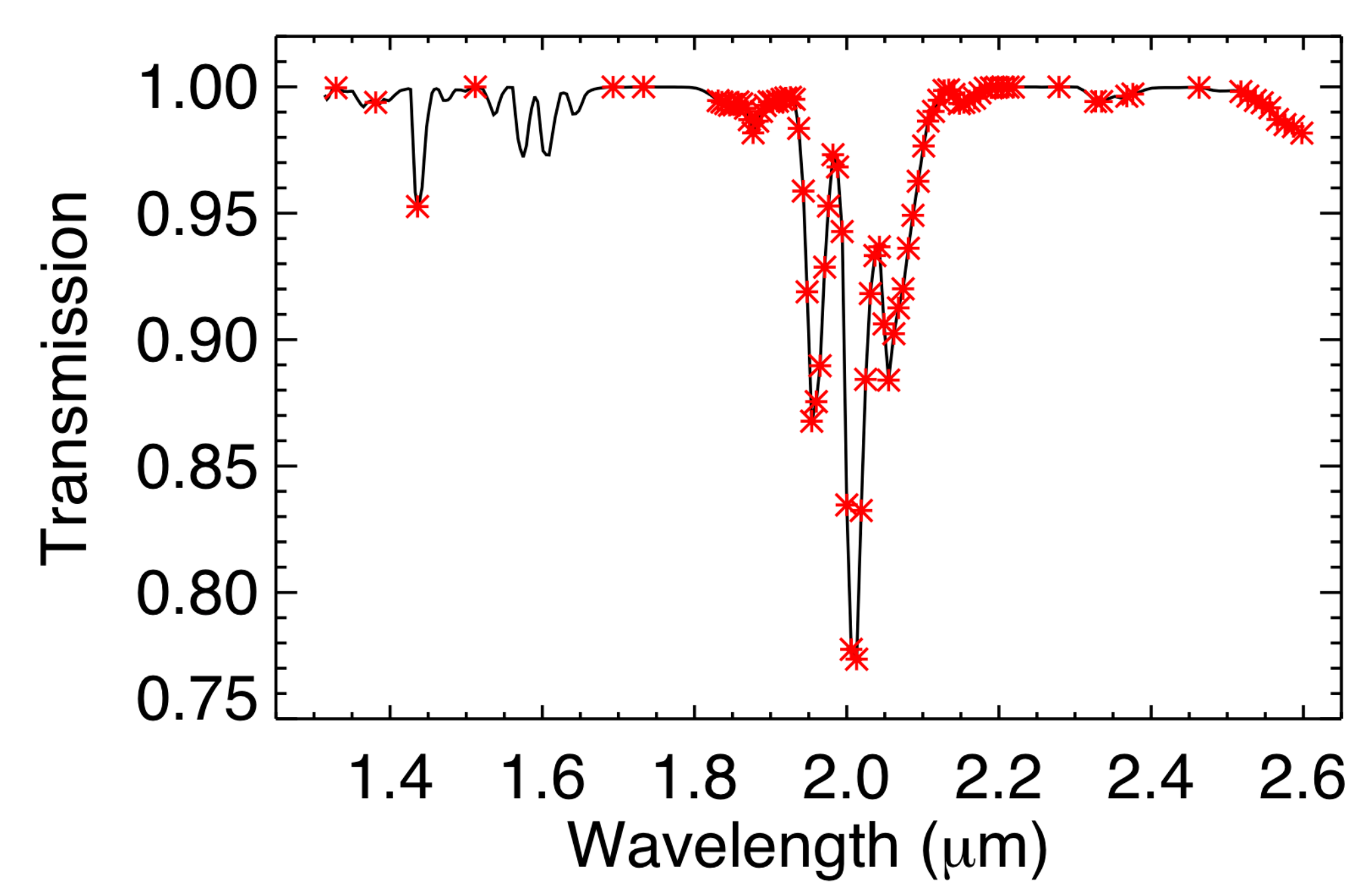}
                    \includegraphics[width=0.85\linewidth,,clip=true,trim=0mm 80mm 0mm 80mm]{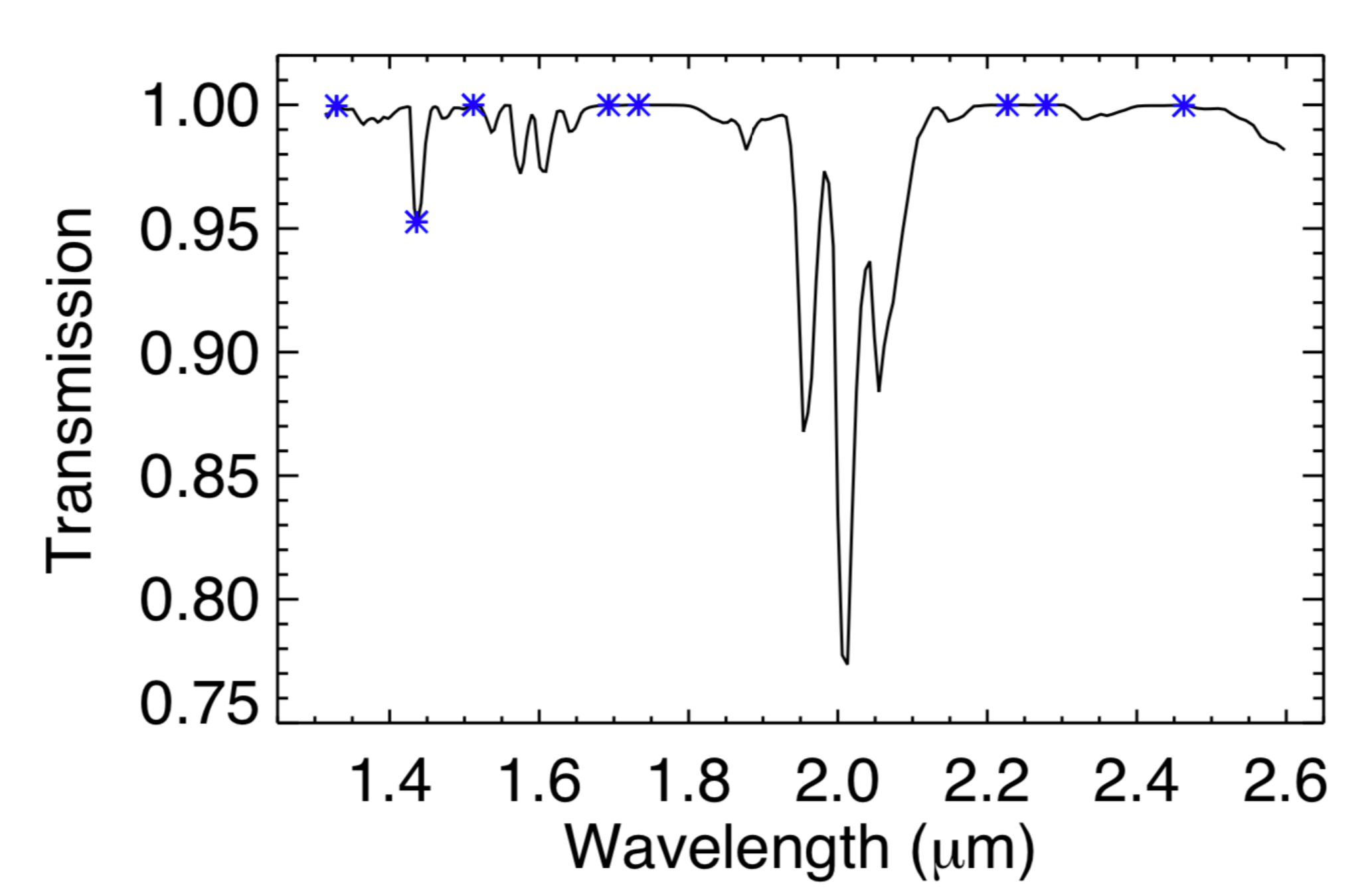}
                    \caption{Wavelenghths sampled Atmosphere Gas Table (upper panel) and the Atmosphere Dust and Ice Table (lower panel) overplotted on Mars atmosphere transmission spectrum.}
                    \label{FigAtmosphereTables}
                \end{figure}

                Figures~\ref{FigNaccTables} and \ref{FigAtmosphereTables} present how the spectrum tables sample some specific mineral spectra or a typical spectrum of atmospheric transmission.

                \begin{table}[]
                    \centering
                    \begin{tabular}{c|c|c}
                        Table ID &  Table nick name & Number of spectels \\
                        \hline
                        \hline
                        Table \#0 & The nominal & 86 \\
                        Table \#1 & The preferred & 100 \\
                        Table \#2 & Odd spectels & 128 \\
                        Table \#3 & Even spectels & 128 \\
                        Table \#4 & All spectels & 256 \\
                        Table \#5 & BD2p50 & 3 \\
                        Table \#6 & BD2p3 & 5 \\
                        Table \#7 & BD2p30 & 3 \\
                        Table \#8 & BD2p34 & 3 \\
                        Table \#9 & BD1p9 Ligth & 3 \\
                        Table \#10 & BD1p9 Full & 6 \\
                        Table \#11 & BD1p4\_1p5 Light & 3 \\
                        Table \#12 & BD1p4\_1p5 Full & 6 \\
                        Table \#13 & Full H$_{2}$O-OH & 19 \\
                        Table \#14 & Atmosphere Dust and ice & 8 \\
                        Table \#15 & Atmosphere gas & 86
                    \end{tabular}
                    \caption{The 16 spectrum tables}
                    \label{NaccTable}
                \end{table}
            
                During Mars operations, the number of successive accumulations will need to be modified, to adapt the total integration time to the actual illumination conditions on the scientific target, using the $\Delta N_\text{acc}$ parameter. This parameter can be set negative or positive. This parameter will uniformly shift the accumulation number of all observed pixels, while maintaining at least one (AOTF ON, AOTF OFF) pair for each observed pixel. Its effect is summarized in the following formula:
                \begin{equation}
                    \text{Actual } N_\text{acc} = 
                    \begin{cases}
                    0& \text{if} N_\text{acc}=0,\\
                    \max(1, N_\text{acc}+\Delta N_\text{acc})& \text{if} N_\text{acc}\geqslant 0
                    \end{cases}
                    \label{EqDeltaNacc}
                \end{equation}
                
            \subsubsection{Auto-exposure}

                The exposure time should be set at a value that excludes any numerical saturation of the ADC converter. As environmental contributions dominate the detector readout current over the observation signal contribution (see Sec.\ref{SecDetection}), the most suitable exposure time strongly depends on the IRS environment, especially its thermal environment, which can only be accurately known after the observations have been scheduled and performed. For this reason, even if any exposure time can be specified between 1 and 255~ms, an auto-exposure function has been implemented to make sure the total readout current lies between 1/3 and 1/2 of the ADC numerical saturation.
            
\begin{figure*}
                    \centering
                    \includegraphics[scale=0.45]{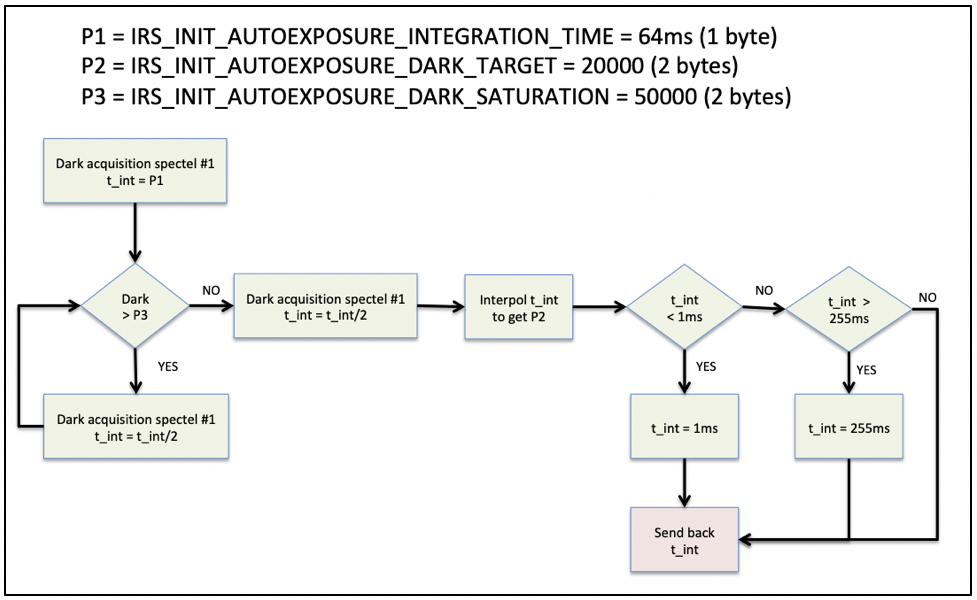}
                    \caption{Decision tree for the determination of the exposure in the auto-exposure mode.}
                    \label{FigAutoexposure}
                \end{figure*}

                Figure~\ref{FigAutoexposure} presents the auto-exposure algorithm that sets the most suitable exposure time.
            
        \subsection{Instrument constraints}
        
            This section describes some operational constraints on the acquisition of an IRS spectrum, arising from the need for certain environmental conditions to be met to ensure the quality of the spectrum and the safety of the instrument.
            
            \subsubsection{TEC ramp}
            
                The TEC cooling down or warming up phases are driven by the FPGA and follow the following rules:
                \begin{itemize}
                    \item The cooling down must not be faster than $-0.5$\textdegree{C}/s until the TEC cold face temperature reaches the TEC setpoint;
                    \item The warming up must not be larger than 1\textdegree{C}/s until the TEC cold face temperature reaches $-25$\textdegree{C}.
                \end{itemize}

                For each change of the TEC setpoint value, the temperature ramp starts back from the current detector temperature value.
                
                A flag, \textit{TECready}, sent in the datareply, gives the status of the detector cooling. The rules determining the value of the \textit{TECready} flag are the following:
                \begin{itemize}
                    \item \textit{TECready}$=1$ for a detector temperature $\leqslant \mathrm{setpoint} +1 $\textdegree{C};
                    \item \textit{TECready}$=0$ for a detector temperature $\geqslant \mathrm{setpoint} +1 $\textdegree{C}.
                \end{itemize}
                
                The status of the \textit{TECready} flag does not forbid spectrum acquisition. It is implemented to convey the information on the TEC status in the data analysis process. 
                
            \subsubsection{Time-out}
            \label{SecTimeOut}
            
                The total duration of a spectrum acquisition, $t_{tot}$, is given by the following formula:
                \begin{equation}
                    t_\text{tot}=2 \cdot \left( t_{int} + 5ms \right) \cdot \left( 1+ \sum_{table} N_\text{acc} + \Delta N_\text{acc} \times N_\text{spectels} \right)
                    \label{EqTotalExposure}
                \end{equation}
                where $t_{int}$ is the spectel integration time specified in the command or established by the auto-exposure procedure, $N_\text{acc}$ the number of accumulations specified for each spectel in the specified $table$, $N_\text{spectels}$ the number of read spectels in the specified table, $\Delta N_\text{acc}$ the specified offset in accumulation applying to all spectels. The additional 5~ms for each spectel are due to the VCO setting time.
                
                When the auto-exposure parameter is set, the total integration time of a single spectrum can not be known in advance. For operation on Mars, this situation would have complicated the schedule and planning of SuperCam and other Perseverance instrument observations. Hence, in order to keep the total acquisition time shorter than a maximum duration, the following time-out process is implemented in the SCMU DPU:
                \begin{itemize}
                    \item A $t_\text{time-out}$ is defined at 150s (this value is stored in the MRAM and can be modified if needed);
                    \item The SCMU DPU calculates $t_\text{tot}$ using Eq.~\ref{EqTotalExposure}. If $t_\text{tot} > t_\text{time-out}$, a $\Delta N_\text{acc}$ is calculated to ensure $t_\text{tot} < t_\text{time-out}$;
                    \item The new accumulation numbers follow the rule set in Eq.~\ref{EqDeltaNacc};
                    \item The new $t_\text{tot}$ is calculated with the new accumulation numbers;
                    \item if $t_\text{tot}$ is still larger than $t_\text{time-out}$, the command is aborted, and the SCMU DPU sends back 0xFF for all spectels in the datareply.
                \end{itemize}
                
            \subsubsection{Environmental constraints}
            
                The instrument can be safely operated in the $\left[ -50, +40 \right]$\textdegree{C} range. However, it will achieve correct SNR, and hence was calibrated (see Sec.~\ref{SecCalibration}), in the $\left[ -35, -5 \right]$\textdegree{C} range.
                
        \subsection{Calibration targets dedicated to the IRS functions}
        \label{SecIRSCalTarget}
        
            SuperCam calibration targets are mounted on the rover. Within the whole set of calibration targets, two are dedicated to the IRS functions: a white target surrounded by dark paint on mechanical holdings, and a dark target. \citet{Manrique2020,Cousin2020} describe specifically and extensively the SuperCam Calibration targets. Here, we will only present information relevant to the two calibration targets devoted to the SuperCam infrared spectrometer calibration and characterization.
            
            The white target purpose is three-fold: i) its spectrum will be used as a standard to determine the relative reflectance of the surface target, ii) it will be observed regularly on Mars to monitor the IRS photometric response and its possible evolution, iii) it will provide a check of the spectral registration by monitoring the sharp atmospheric CO$_2$ absorptions. The black paint around the white target will provide the highest possible radiometric contrast to monitor the evolution of the IRS PSF and alignment with respect to the RMI boresight by scanning vertically or horizontally across the white target. Finally, the black target will be used to evaluate the possible contribution of the far side lobes of the point spread function (PSF) of the instrument.
            
            The white target is made of cintered aluminum oxide provided in the commercial name of AluWhite 98 by Avian Technologies. It exhibits hemispherical reflectance in excess of 98\% in the IRS spectral range. The matte finish was preferred to the glossy finishing because of its near-Lambertian properties. AluWhite 98 was preferred over other solutions, such as Spectralon, for its greater resistance to UV radiation expected on Mars. The black target is painted with Aeroglaze paint Z307, chosen for its high resistance to the harsh space environments. The two IRS calibration targets have a visible diameter of 8 mm and are located at 1.56~m away from the SuperCam entrance window. The angle between the normal to the calibration targets and the Perseverance $-Z$ axis (hence the zenith when the Rover is horizontal), is 50$^{\circ}$, and the angle between the normal to calibration targets and the boresight of the SuperCam telescope (i.e. the emergence angle) is $13^{\circ}$. In order to minimize dust deposition on these targets, a 11-mm magnet is mounted behind each of them to keep the central part of the target clear of dust over a diameter of 4~mm. This clean zone is twice the size of the 1.15-mrad FOV at 1.56~m.
            
            We performed a broad laboratory characterization of the two IRS calibration targets. This characterization was performed with different facilities at two different institutions: the Université Grenoble Alpes (France) with a Spectro-goniometer thoroughly described in \citet{Potin2018}, and at the University of Bern (Switzerland) with a goniometer presented in \citet{Pommerol2011}. In Grenoble, the reflectance was sampled between 600 and 3000~nm at a sampling step size of 100~nm. The emergence angle was fixed to 12$^{\circ}$ while variations in  azimuth angle was explored at 0$^{\circ}$, 10$^{\circ}$, 20$^{\circ}$, 40$^{\circ}$, 60$^{\circ}$, 80$^{\circ}$, 100$^{\circ}$, 120$^{\circ}$, 140$^{\circ}$, 160$^{\circ}$, 170$^{\circ}$, and 180$^{\circ}$, and for the incidence angle at 5$^{\circ}$, 20$^{\circ}$, 30$^{\circ}$, 40$^{\circ}$, 50$^{\circ}$, and 60$^{\circ}$. In Bern, as described in \citet{Kinch2020}, the reflectance was measured with six color filters at 450, 550, 650, 750, 910, and 1064~nm. The phase function was explored at emergence angles in the range 0$^{\circ}$--80$^{\circ}$ with 5$^{\circ}$ steps, azimuth angles in the range 0--180 at 15$^{\circ}$ steps, and for the following incidence angles: 0$^{\circ}$, 15$^{\circ}$, 30$^{\circ}$, 45$^{\circ}$, 58$^{\circ}$, and 70$^{\circ}$.
            
\begin{figure*}
                \centering
                \includegraphics[width=10cm,angle=270]{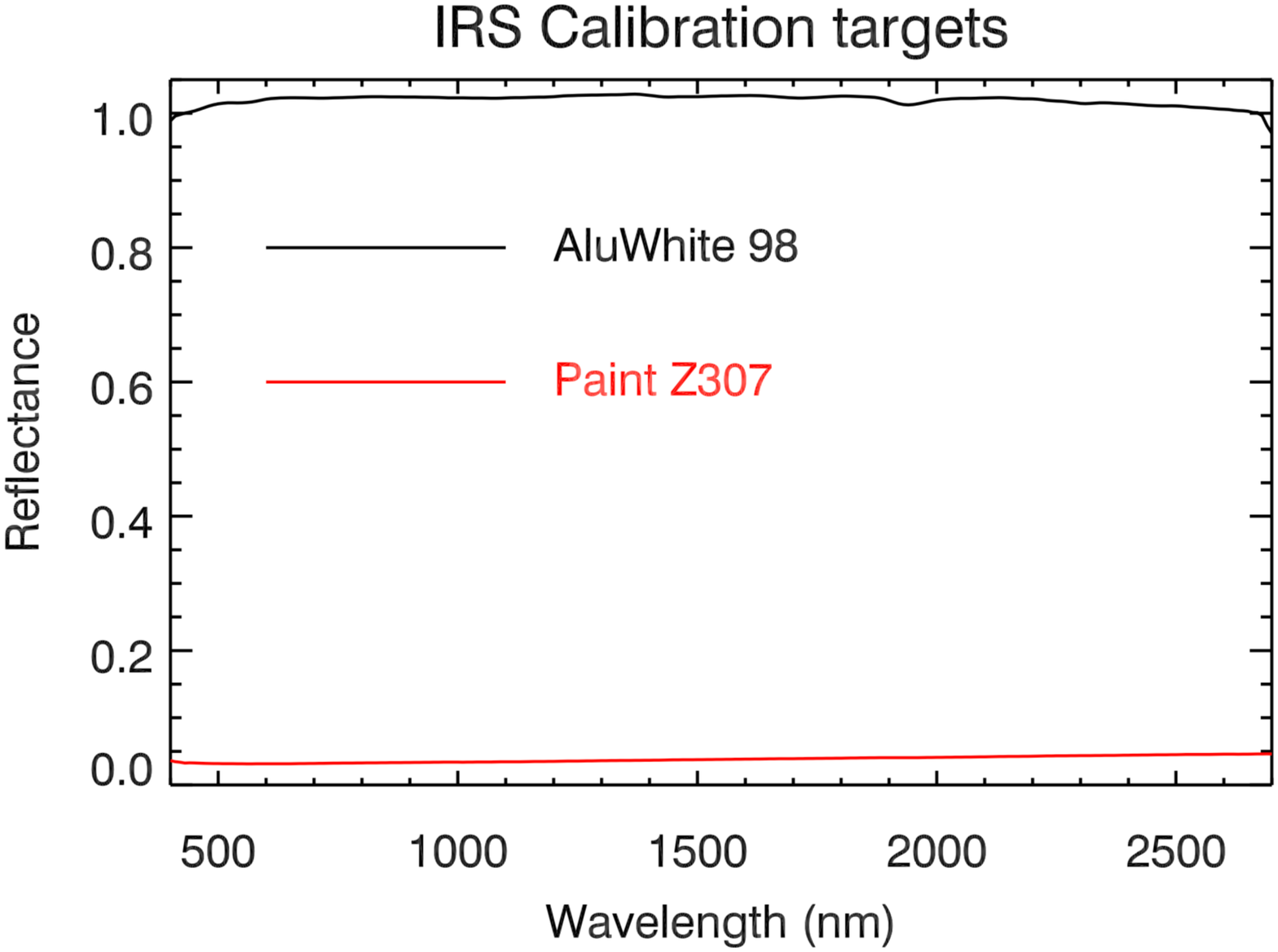}
                \includegraphics[width=10cm, clip=true, trim=35mm 60mm 100mm 40mm]{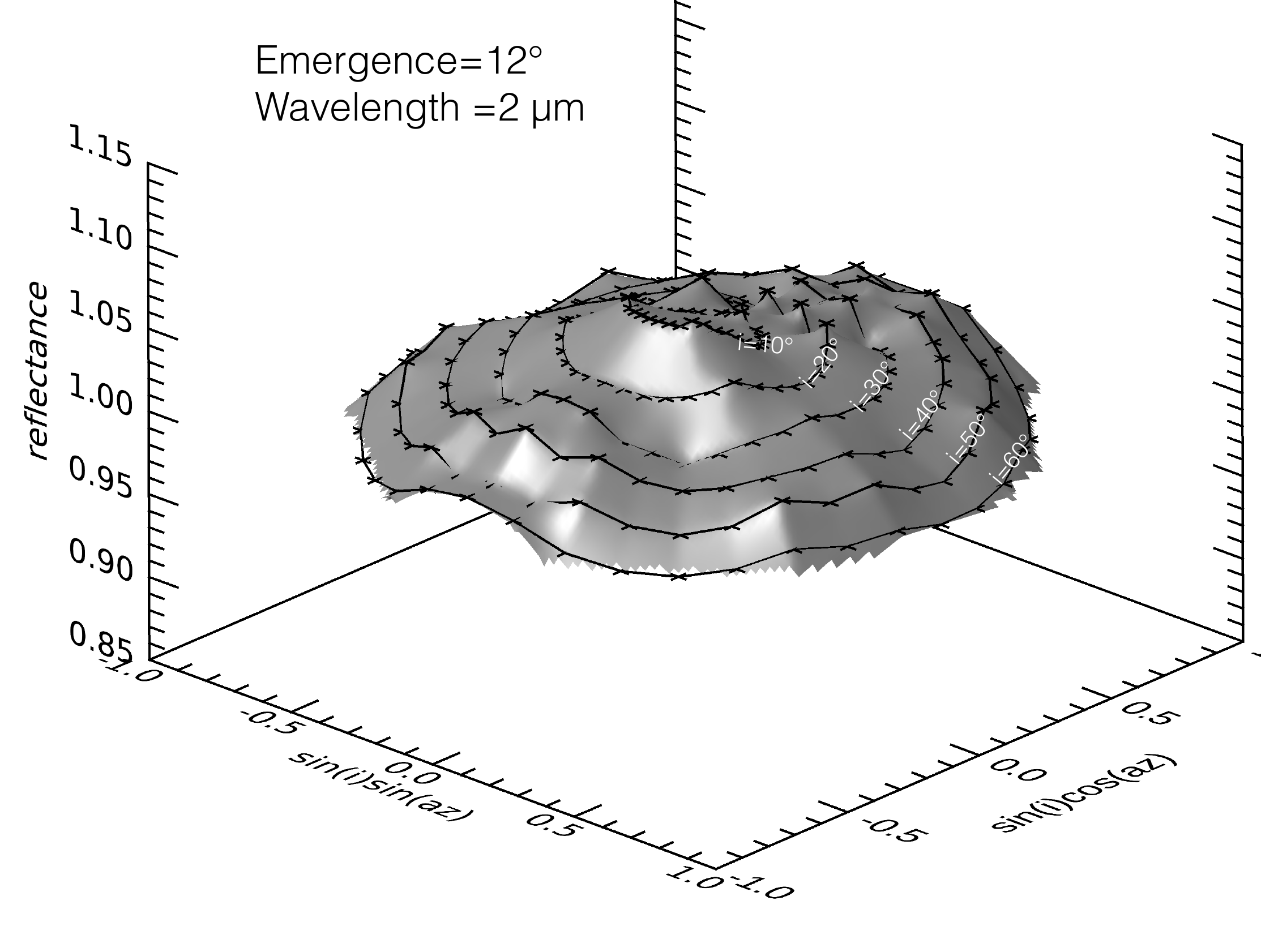}
                \caption{Upper panel: Reflectance spectra for the AluWhite 98 (black line) and Paint Z307 (red line) registered at Université Grenoble Alpes for an incidence angle of 30$^{\circ}$ and an emergence angle of 0$^{\circ}$. Lower panel: BRDF of the AluWhite target obtained at Université Grenoble Alpes at 2~$\mu$m for an emergence angle of 12$^{\circ}$.}
                \label{FigCalTargets}
            \end{figure*}

            The results are summarized in Fig.~\ref{FigCalTargets} that displays the white and black targets reflectances (left panel) throughout the spectral range for incidence and emergence angles of $30^{\circ}$ and $0^{\circ}$ respectively, and the phase function mapping (right panel). The error on the reflectance is estimated to be smaller than 0.01. The reflectance spectral variations are smaller than 1\%, which is fully compliant with our requirement for a relative calibration better than 1\%. The white target also exhibits excellent Lambertian behaviour, with variations not in excess of 10\%, well within our goal of a 20\% absolute radiometric calibration. The dark Z307 paint also is extremely spectrally smooth, with reflectance rising gently from 0.035 to 0.045 within our spectral range.
            
    \section{Instrument testing, calibration and performance}
    \label{InstrumentTest}
    
        The SuperCam infrared spectrometer was fully tested, validated and calibrated, first at the IRS stand-alone level, and finally at the SCAM-MU level for radiometric calibration. These tests were conducted in the relevant thermal and environmental conditions using the LESIA facilities. We also present in Sec.~\ref{SecMarsCalibration} how we project to routinely check the IRS performance while on operations in Jezero crater.
    
        \subsection{Spectral performance}
        \label{SecSpectralPerformances}
        
\begin{figure}
            \centering
            \includegraphics[scale=0.4, clip=true, trim=8mm 0mm 200mm 0mm]{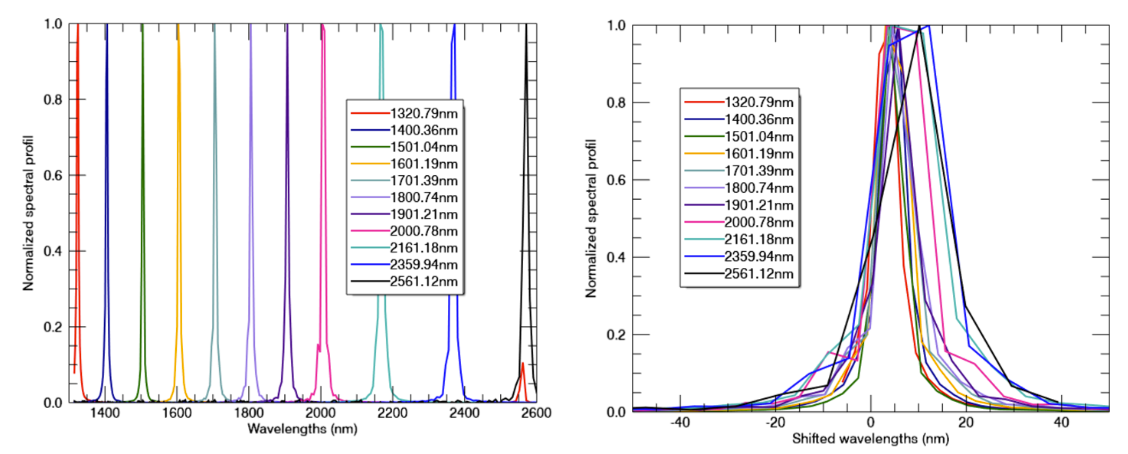}
            \caption{Normalized IRS responses to the monochromator at different wavelengths.}
            \label{Monochrom}
        \end{figure}

\begin{figure*}
            \centering
            \includegraphics[scale=0.4]{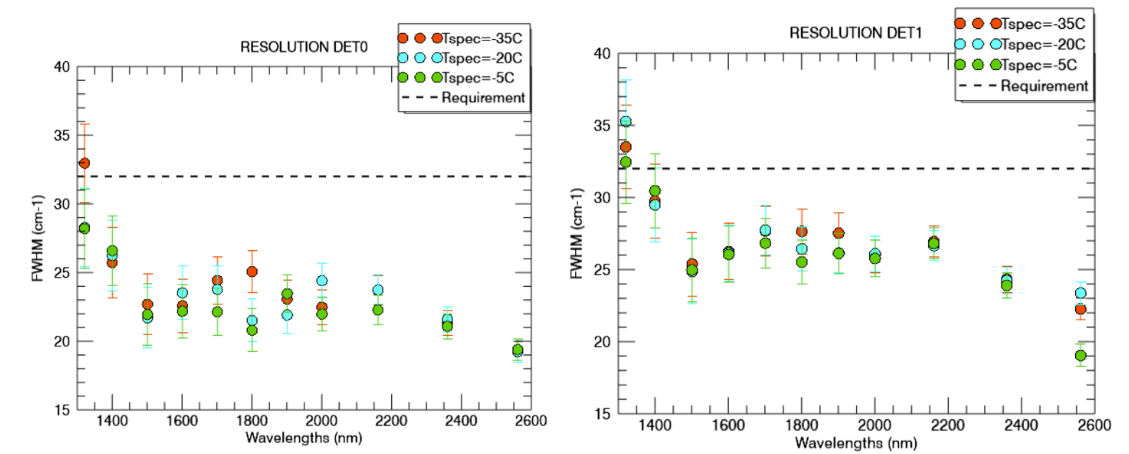}
            \caption{Measured spectral resolution as a function of wavelength for the three tested temperatures, for the nominal detector (left panel) and the redundant detector (right panel).}
            \label{Resolution}
        \end{figure*}

            The IRS spectral resolution, the spectral registration, and the spectral bandpass were measured using a calibrated monochromator, a calibrated Fabry-Perot and the stand-alone IRS at 3 different temperatures: $-35$\textdegree{C}, $-20$\textdegree{C} and $-5$\textdegree{C}. A gaussian fit was applied to the IRS response signal. The central wavelength of the gaussian fit provided the spectral registration, while the FWHM of the gaussian fit, deconvolved by the monochromator intrisic width provided the IRS spectral resolution. Figure~\ref{Monochrom} shows the IRS response to the monochromator input at some selected wavelengths, while Fig.~\ref{Resolution} displays the spectral resolution measured for both detector optical paths at the three different temperatures. The full spectral bandpass was measured to cover precisely the spectral range [1.3148 -- 2.599]~$\mu$m.
            
            The spectral registration was found to be hardly dependent on the AOTF temperature: at 1.6~$\mu$m, it varies by 3~nm over the temperature range. The same is true for the spectral resolution, which varies by less than 1\% across the temperature range.
            
            The exact line spread function of the AOTF is not a gaussian, but features side lobes. The amplitude of the side lobes was specified and tested by the manufacturer G\&H to be smaller than -13~dB of the main lobe.
        
\begin{figure*}
            \centering
            \includegraphics[scale=0.4]{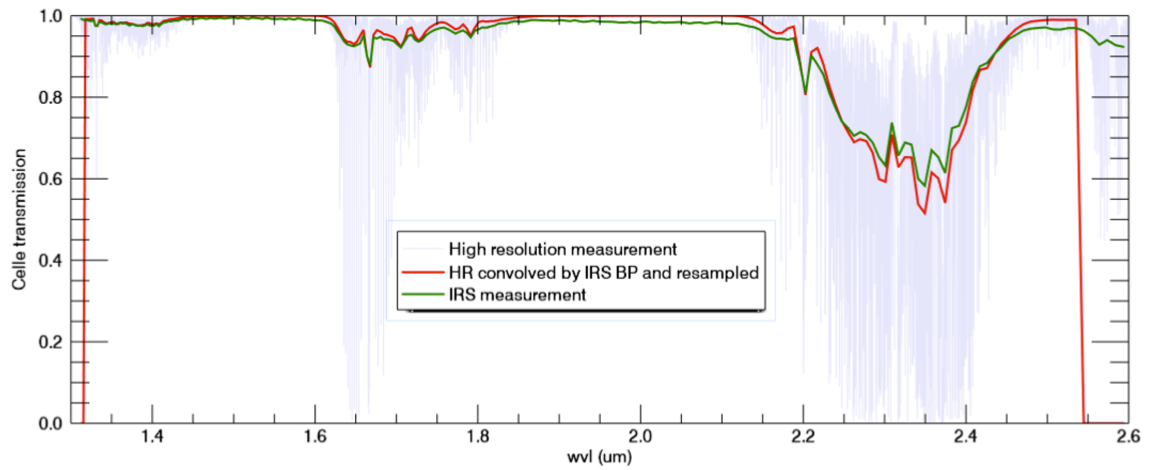}
            \caption{Transmission spectrum of a CH$_4$ gas cell measured by the IRS (green line) compared with a synthetic CH$_4$ spectrum  (red line).}
            \label{CH4}
        \end{figure*}

        As a cross check of the spectral registration and spectral resolution, we measured the transmission of a CH$_4$ gas cell with the IRS. Illuminated by a halogen source, the CH$_4$ gas cell transmission was first measured at high spectral resolution, then convolved by the IRS line spread function and re-sampled on the IRS registration and bandpass. The line spread function was taken as constant over the full bandpass. The correlation between both spectra is better than 0.99, as shown in Fig.~\ref{CH4}.
        
        \subsection{Radiometric performance}
        
            The noise level was evaluated using the Flight Model (FM) of the IRBOX but only an Engineering Model (EM) of the IRBOARD, both connected by a long harness. This configuration was not optimal, as the EM IRBoard had lower performance compared to the FM IRBoard, and as the long harness contributed some additional noise. Hence, our setup yielded only an upper limit of the noise level, comprising contributions from the IRS environment, photodiode readout and the FEE.
            
\begin{figure*}
                \centering
                \includegraphics[scale=0.4]{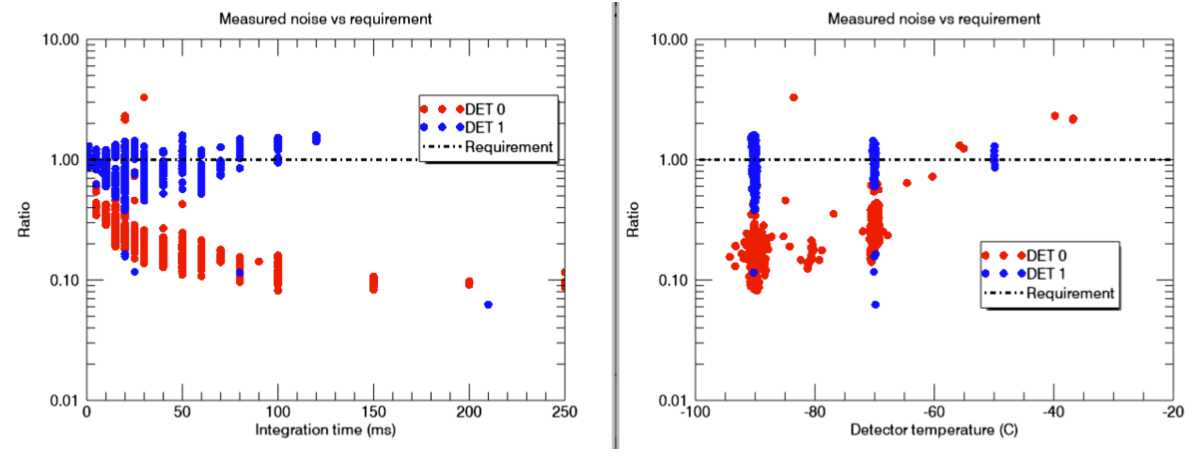}
                \caption{Noise measurement, normalized to the level L5 requirement of SNR$=56$, as a function of integration time (left panel) and the detector temperature (right panel) for all configurations, for detector \#0 (red points) and detector \#1 (blue points). The L5 requirement (Tab.~\ref{TabRequirementL5} requirement) is shown as the dotted line.}
                \label{noise}
            \end{figure*}

            The measurements were performed at various accumulation numbers (Nacc), integration times, and RF power, and were normalized to the L5 requirement parameters (acquisition of 86 spectels in a total integration time of 80~seconds at Mars mean solar irradiance, see Tab.~\ref{TabRequirementL5}). To be consistent with our acquisition strategy, the noise level was evaluated on a spectrum obtained by the subtration of the AOTF ON and AOTF OFF signals. Figure~\ref{noise} displays a scatter plot of all these noise measurements (for all instrument configurations in terms of instrument and detector temperatures, integration times, number of accumulation,...) relative to the requirement as a function of the integration time or the detector temperature. This demonstrates that for the photodiode \#0, the requirement is met whatever the chosen configuration. Photodiode \#1 is noisier, and the requirement is not met for some configurations.
            
        \subsection{Relative and absolute radiometric calibration}
        \label{SecCalibration}
        
            The relative and absolute radiometric calibrations campaign is thoroughly described in a dedicated paper \citep{Royer2020}. We present here only a brief outline of the calibration strategy, implementation and results.
        
            The L5 requirements (Tab.~\ref{TabRequirementL5}) target an absolute radiometric calibration accuracy better than $20\%$, and a relative radiometric calibration better than $1\%$ up to 2.55~$\mu$m, and better that $2\%$ longward of 2.55$\mu$m. This later difference in requirements across the wavelength range arises from the difficulty of drying and stabilizing the humidity of the environment in the testing facility. Hence, the same accuracy could not be achieved within and outside of the strong 2.6-$\mu$m atmospheric water band.
        
            The IRS, fully integrated in the MastUnit to be representative of the whole optical path, was integrated in a thermal and vacuum chamber (SimEnOn) on LESIA premises. The Mast Unit was coupled to a cold plate for thermal regulation. The optical stimulus was provides by a tunable black body purchased to CI Systems (Ref.~SR-200-32). A tube, constantly flushed with dry nitrogen to limit water vapour absorptions, was designed by the Institut d'Astrophysique Spatiale (IAS) to link the black body to the SimEnOn entrance window, manufactured using saphire for infrared transparency.
        
            The two photodiodes were first tested for linearity in charge by scanning the integration time from 1 to 255~ms, and the linearity in flux by varying the blackbody temperature from 200$^{\circ}$C to 600$^{\circ}$C. Due to time constraints, only 9 and 2 spectels were tested respectively for linearity in charge and linearity in flux. The two photodiodes were linear in charge within $0.1\%$ compliant with the need for a 1\%-relative calibration. They were measured as slightly non-linear in flux, with the actual behaviour being a power law detailed in \citet{Royer2020}.
        
            The radiometric calibration was performed using three different blackbody temperatures (180, 450 and 600$^{\circ}$C, with an uncertainty of $\pm2.5^{\circ}$C), representative of the expected fluxes across the entire the 1.3--2.6~$\mu$m range for nominal illumination conditions at the surface of Mars. Due to time constraints, it was not possible to test all the possible operational configurations and to cover the whole phase space of environmental parameters. The priority was to calibrate more extensively the nominal photodiode \#0 than the redundant photodiode \#1. The lowest environmental temperatures, for which better signal-to-noise ratio will be achieved, were given the priority over warmer temperatures. We explored the range $[-35^{\circ},-5^{\circ}]$C for the spectrometer temperature, and the range $[-90^{\circ},-70^{\circ}]$ for the photodiode \#0 cold temperature.
        
            From these measurements, the Instrument transfer function was estimated using the equation
            \begin{equation*}
               \mathrm{Signal} - \mathrm{Dark}= ITF_\mathrm{fac} \times t_{\mathrm{int}} \times \phi^{ITF_\mathrm{exp}}
            \end{equation*}
            where $t_{\mathrm{int}}$ and $\phi$ are the integration time and the black body flux, and $ITF_\mathrm{fac}$ and $ITF_\mathrm{exp}$ two parameters depending on the environmental conditions. These two parameters and their dependencies were retrieved using a Monte-Carlo by Markov Chain (MCMC) method to fit all the experimental points obtained. The derived instrument transfer function is presented and discussed in \citet{Royer2020} and implemented in the IRS calibration pipeline.

        \subsection{Data processing by the IRS pipeline}
        \label{SecPipeline}
        
\begin{figure}
                \centering
                \includegraphics[scale=.25]{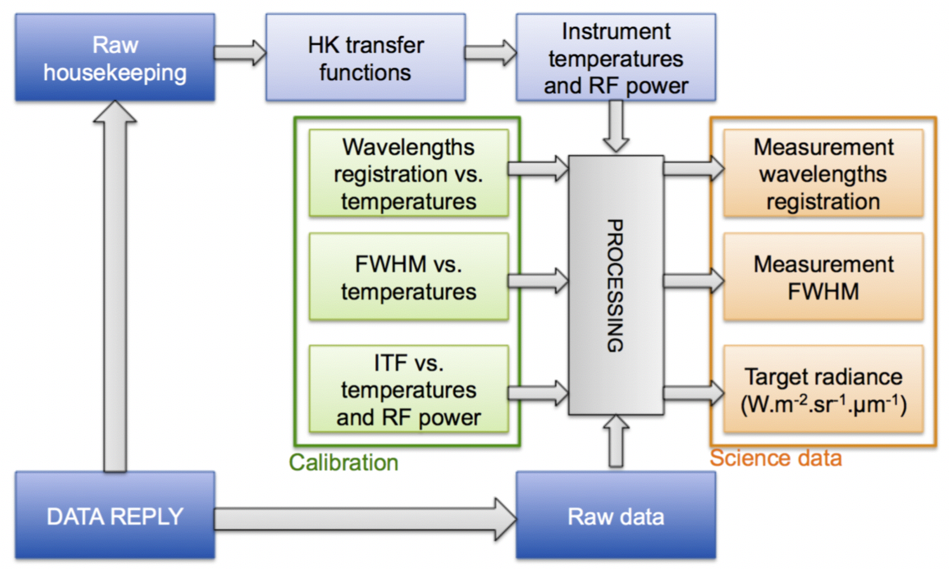}
                \caption{Data processing principle}
                \label{dataprocessing}
            \end{figure}

            Figure~\ref{dataprocessing} presents the flowchart describing how the data reply transmitted by the rover is handled by the IRS pipeline to generate the scientific data, using the housekeeping measurements.

            The IRS pipeline is designed to process raw data files from the IRS Flight Model (FM) but also from the IRS Engineering Qualification Model (EQM) during pre-flight testing, combining raw target and dark spectra to obtain:
            \begin{itemize}
                \item a calibrated spectrum in radiance ($\text{W/m$^2$/sr/$\mu$m}$) with calibrated wavelengths,
                \item a calibrated spectrum corrected for a standard Mars atmospheric transmission,
                \item an I/F spectrum: calibrated spectrum divided by the solar irradiance (only for FM).
            \end{itemize}

            The IRS pipeline also processes the IRS HK of the data reply to provide the user with the key instrumental parameters (see~\ref{TabHK}).

            The first step consists in initialization and extraction of all the relevant parameters from the data reply. Specific geometric parameters (solar zenith angle,  solar incident angle on the calibration targets, ...) are provided by JPL calculations using the rover kernels.

            The second step consists of spectral calibration. The wavelength of each spectel depends on the AOTF RF power and temperature at the beginning of the measurement. The output spectel wavelengths are obtained by interpolating the spectral registration measured during the pre-flight tests (see Sec.~\ref{SecSpectralPerformances}) as a function of RF power and temperature fitted with a 4-degree polynomial. The pipeline applies the same process to calculate the FWHM of each spectel. 

            The third step is the radiometric conversion from DN to radiance using the instrument transfer function (ITF) obtained on ground by \citet{Royer2020}. First, the raw target and the raw dark spectra, in DN, are divided by the Nacc Table value for each spectel. At this point, the saturation is checked. If the signal or the dark is greater than 60000 DN, the spectel is flagged as saturated and rejected. Next, the dark spectrum is subtracted from the target spectrum, the ITF is interpolated in the spectrometer-temperature and photodiode-temperature space, and the radiance is retrieved in $\text{W/m$^2$/sr/$\mu$m}$ using Eq.~3 of \citet{Royer2020}.

            As another data product, the IRS pipeline divides the radiance spectrum by a Mars atmospheric transmission. The Mars atmospheric transmission was computed for the annual mean Jezero surface pressure (see Fig.~\ref{FigClimatologie}) for solar zenith incidence angles between 0° and 60°, interpolated for the true solar zenith angle. No dust scattering is included in the Mars atmospheric transmission. The final data product, I/F, is then obtained by dividing by the solar irradiance at the time of observation.

        \subsection{Operation on Mars}

            As the acquisition duration of a single spectrum will range between 1 and 3 minutes, we do not expect to acquire more than a few 10s of spectra per sol on Mars.
            
            In principle, near-infrared relative reflectance spectra should be ratioing two successive observations of the surface target and the white calibration target acquired by the same solar elevation and azimuth angles. This observation strategy can precisely account for the atmospheric absorptions and for the relative contributions of the direct solar illumination and the sky diffuse illumination due to dust and cloud scattering.  Visible and near-infrared measurements on numerous Mars landed missions (Viking through MSL) have used similar calibration targets as a standard for calibration of on-sensor radiance measurements to relative reflectance.  However, one challenge confronting this method is that the solar incidence and phase angles will differ between the IRS calibration target and on the surface target. The surface targets will most frequently lie in or near the rover work space, where the nearby surface field of view is best, and often to be eventually analyzed by the Robotic Arm, while the calibration target is located on the deck at the rear of the rover. As a consequence, the relative contribution of direct and diffusive illumination will differ between the surface and calibration targets, limiting the precision of the atmospheric correction. A second challenge is that the white calibration target must be sunlit and observed near in time to the surface observation to provide similar solar illumination and atmospheric conditions. This may not always be possible depending on rover orientation relative to the Sun, particularly given the inclined mounting of the SCCT. Fortunately, the near-Lambertian nature of the Aluwhite98 calibration target will provide a consistent radiometric response under a variety of illumination conditions. During rover operations when the calibration target is not sunlit, a measurement taken on a subsequent or previous sol at or near the same time of day under similar atmospheric opacity will be used, as has often been done for previous landed mission visible/near-infrared calibrations. For measurements acquired near the rover, the difference in atmospheric path lengths between the rover and surface target versus the SCCT are minimal and should not greatly impact the resulting ratio spectra. By contrast, for long distance observations ($>$ 10s m), the I/F method described in Section~\ref{SecPipeline} may provide better results for longer path lengths.  However, the exclusion of dust opacity in the atmospheric transmission models may introduce complications owing to the seasonally and diurnally variable atmospheric dust opacities on Mars. As such, both techniques will be investigated under a variety of observing conditions to determine which method is preferred over the wavelength ranges sampled by both the IRS and Body Unit spectrometers. 

    \subsection{Calibration on Mars}
    \label{SecMarsCalibration}
    
        The SuperCam calibration targets will be used to check possible evolution of the IRS transfer function and tuning relations throughout the whole Mars2020 mission. They will also be observed over the course of the mission to provide illumination reference spectra that will be used for comparison of reflectance spectra to surface targets.
        
        For this purpose, we plan to obtain, as early as possible after landing, a full set of white target observations while it is as clean as possible, i.e. before non-magnetic dust deposition obscures the surface of even the areas encompassed by the magnets (Sec.~\ref{SecIRSCalTarget}). Over the course of the mission we plan to observe regularly the white and black targets to check the instrument radiometric calibration together with the PSF form and centering. Given the known radiation environment at the surface of Mars, given that SuperCam will be heated during Martian nighttime to reduce the thermal cycling, we do not expect rapid ageing of the mirrors, dichroics, and photodiodes. The AOTF transmission  could vary because of an ageing electronics or because of a less efficient  transducer - crystal coupling. Our plan is to verify the radiometric response on a monthly basis.
        
        The spectral registration will be also checked by investigating the 2.0-$\mu$m CO$_2$ band position, as well as through observations of targets with pronounced spectral bands (such as the white paint used on the SCCT housing). Given the high signal-to-noise ratio that will be achieved on a white calibration target spectrum, it will be possible to verify the spectral registration to a fraction of a FWHM. To quantify this precision, we generated an Mars atmosphere transmission spectrum using a line-by-line radiative transfer code \citep{Fouchet2007}, convolved it with the IRS line spread function, added some gaussian random noise, sampled it at wavelength shifted by a small value $\Delta\lambda$  from the expected wavelengths. For wavelength shifts larger than $\Delta\lambda=0.1$~ nm, the mismatch between the shifted spectrum and nominal spectrum becomes evident. This procedure will allow for a check for the spectral registration at the surface of Mars compliant with the IRS L5 requirements.
       
\section{Conclusions}

    The design and implementation of the SuperCam infrared spectrometer had to overcome several challenges and obstacles. Indeed, SuperCam is inherited from the ChemCam instrument, successfully operating from the surface of Mars on board the Curiosity since 2012. Given this success, a radical modification of the ChemCam design was unwarranted. Rather, inclusion of the infrared spectrometer was designed to minimize the changes on the mechanical, electronical and optical systems from ChemCam to SuperCam.
    
    In this context, it was not possible to accommodate the infrared spectrometer inside the Body Unit, as the 6-m long optical fiber between the Mast Unit and Body Unit would have introduced spurious OH signatures that would have considerably reduced the detection sensitivity and accuracy of the IRS on some minerals key to achieve the scientific objectives of the Mars 2020 mission. Instead, the infrared spectrometer had to be accommodated inside the Mast Unit, at the expense of its volume.
    
    As the Mast Unit is insulated inside the Remote Warm Electronics Box (RWEB) to easily maintain all the systems within their non-operational allowable flight temperatures during night time, it is difficult to dissipate the heat usually generated by cooled large infrared detector. For this reason, 2D matrix detectors that would have allowed acquisition of hypersprectal images were impossible to accommodate, and the design had to rely on a photodiode and an AOTF dispersing system.
    
    Finally, SuperCam is an instrument that required collection of light from 245 to 2600~nm, and collected light must be transmitted to the Body Unit spectrometers, the auto-focus photodiode, the RMI, and the IRS, while excluding two laser wavelengths at 532 and 1064~nm. Surface coating of the optical pieces were thus optimized for a broad wavelength range rather than just the infrared range, while several dichroics split the light beam to redirect it towards the various subsystems. It resulted from this complexity that the SuperCam infrared throughput is smaller than the usual throughput of dedicated infrared spectrometers, which induced strong constraints on the detection system and its electronics. 

    This paper presented how the technical design of the SuperCam infrared spectrometer overcame these hurdles and difficulties to implement an instrument in line with the overarching scientific goals of the Mars2020 mission. Although a stand-alone infrared spectrometer may have enabled higher intrinsic IR performance, we do believe that the co-aligned and nested LIBS, Raman, and IR observations of a single target will overall provide more precise and in-depth characterization of the Jezero geology and mineralogy.

\appendix
\section{Appendix A: Description of housekeeping data}
\label{AppendixHK}

The following probes, and their associated transfer functions, are implemented in the SuperCam infrared spectrometer:
                \begin{itemize}
                    \item Two AD 590 thermisters measure the AOTF and the TEC hot face temperature. Those measurements are written in the datareply in 8 bits and the HK data file in 16 bits with the following transfer functions:
                    \begin{align*}
                        &\text{8 bits:} && T(^\circ C)=0.512 \times T_{ADU}-51.721 \\
                        &\text{16 bits:} && T(^\circ C)=0.001992 \times T_{ADU}-51.721 
                    \end{align*}*
                    \item The Judson-provided TEC sensor measures the TEC cold face temperature;
                    \begin{equation*}
                        T(^\circ C)=-21.70105 \times \log\left(\frac{180000\times T_{ADU}}{2^{16}-1-T_{ADU}}\right)+182.34593
                    \end{equation*}
                    \item The TEC cold side setpoint;
                    \begin{equation*}
                        T(^\circ C)=-21.70105 \times \log\left(\frac{180000\times T_{ADU} \times 256}{2^{16}-1-T_{ADU} \times 256}\right)+182.34593
                    \end{equation*} 
                    \item IRS 3V as the FEE power-voltage
                    \begin{equation*}
                        V_{3V} (V)=\frac{V_{ADU}}{32768 \times 3.33 \times 2}
                    \end{equation*}
                    \item The IRS 20V RF for powering the AOTF driver;
                    \begin{equation*}
                        V_{20V}(V)=\frac{V_{ADU}}{32768 \times 3.33 \times 7.49}
                    \end{equation*}
                    \item IRS $\pm$ 5V RF and $\pm$ 8V DT $\pm$ Voltage powering the AOTF and TEC drivers
                    \begin{equation*}
                        V(V)=\frac{\pm V_{ADU}}{32768 \times 3.33 \times 3} 
                    \end{equation*}
                \end{itemize}



\bibliographystyle{cas-model2-names}

\bibliography{IRS}








\end{document}